\documentclass[usenatbib]{mn2e}
\usepackage{psfig}

\def\gsim{\mathrel{\raise0.35ex\hbox{$\scriptstyle >$}\kern-0.6em
\lower0.40ex\hbox{{$\scriptstyle \sim$}}}}
\def\lsim{\mathrel{\raise0.35ex\hbox{$\scriptstyle <$}\kern-0.6em
\lower0.40ex\hbox{{$\scriptstyle \sim$}}}}
\def\gs{\mathrel{\raise0.35ex\hbox{$\scriptstyle >$}\kern-0.6em
\lower0.40ex\hbox{{$\scriptstyle \sim$}}}}
\def\ls{\mathrel{\raise0.35ex\hbox{$\scriptstyle <$}\kern-0.6em
\lower0.40ex\hbox{{$\scriptstyle \sim$}}}}

\def\kms{\,\hbox{km}\,\hbox{s}^{-1}}

\def\Wm2{\,\hbox{W}\,\hbox{m}^{-2}}

\begin{document}

\title[Resolved Spectroscopy of z=1 Giant Luminous Arcs]{Galaxies Under
  the Cosmic Microscope: \\ Resolved Spectroscopy and New Constraints
  on the \\ z=1 Tully-Fisher relation}

\author[Swinbank et al.]{
\parbox[h]{\textwidth}{
A.\ M.\ Swinbank,$^{1\, *}$
R.\ G.\ Bower,$^1$
Graham P.\ Smith,$^{2,\,3}$
Ian Smail,$^1$\\
J.-P.\ Kneib,$^{4,\,2}$
R.\ S.\ Ellis,$^2$
D.\ P.\ Stark,$^2$
\& A.\ J.\ Bunker$^5$}
\vspace*{6pt} \\
$^1$ Institute for Computational Cosmology, Department of Physics,
     Durham University, South Road, Durham DH1 3LE\\
$^2$ California Institute of Technology, MC 105-24, Pasadena, CA 91125,
     USA \\ 
$^3$ School of Physics and Astronomy, University of Birmingham,
     Edgbaston, Birmingham, B15 2TT\\ 
$^4$ Laboratoire d'Astrophysique de Marseille, Traverse du Siphon -
     B.P.8 13376, Marseille Cedec 12, France\\ 
$^5$ University of Exeter, School of Physics, Stocket Road, Exeter,
     EX4 4QL \\ 
$^*$ Email: a.m.swinbank@dur.ac.uk \\
}
\setcounter{footnote}{0}

\maketitle

\begin{abstract}
  We exploit the gravitational potential of massive cluster lenses to
  probe the emission line properties of six $z=1$ galaxies which appear
  as highly magnified luminous arcs.  Using the GMOS integral field
  spectrograph together with detailed cluster lens models we
  reconstruct the intrinsic morphologies and two-dimensional velocity
  fields in these galaxies on scales corresponds to $\sim$0.5kpc
  (unlensed) at $z=1$.  Four of the galaxies have stable disk-like
  kinematics, whilst the other two resemble interacting or starburst
  galaxies.  These galaxies lie close to the mean rest-frame $I$-band
  Tully-Fisher relation for nearby spirals suggesting a clear
  preference for hierarchical growth of structure.  In the rest-frame
  $B$-band, the observations suggest $0.5\pm0.3$mag of brightening,
  consistent with increased star-formation activity at $z=1$.  However,
  the galaxies with stable disk kinematics have more slowly rising
  rotation curves than expected from galaxies with similar surface
  brightness in the local Universe.  We suggest that this may arise
  because the distant galaxies have lower bulge masses than their local
  counter-parts.  Whilst this study is based on only six galaxies, the
  gain in flux and in spatial resolution achieved via gravitational
  magnification provides a much more detailed view of the high redshift
  Universe than possible with conventional surveys.
\end{abstract}


\section{Introduction}
Massive galaxy clusters magnify the light from galaxies that
serendipitously lie behind them. This natural magnification provides
the opportunity to study intrinsically faint high redshift galaxies
with a spatial resolution and to surface brightness limits that cannot
be attained via conventional observations.  These highly magnified
sources can provide unique insights into the properties of typical
galaxies at early times in the Universe \citep{Smail96, Franx97,
  Teplitz00, Ellis01, Campusano01, Smith02, Swinbank03, Kneib04b}.  The
majority of known galaxy cluster lenses are at $z\lsim0.3$, which makes
them ideal for detailed studies of galaxies at $z\gsim1$ \citep{Edge03,
  Sand05}, i.e.\ before the Universe had reached half of its current
age.

At $z\gsim1$ important questions regarding the relationship between the
total mass of a galaxy and the baryonic mass locked up in stars remain
unanswered.  In local rotationally-supported spiral galaxies, this
relation is best described by the Tully-Fisher relation
\citep{TullyFisher} -- an empirical correlation between the terminal
rotational velocity (or line width) and the absolute magnitude of a
spiral galaxy.  This relationship may reflect how
rotationally-supported galaxies formed, perhaps suggesting the presence
of self regulating processes for star formation in galactic disks.
Measuring the evolution of the Tully-Fisher (TF) relation as a function
of look back time therefore provides important insights into the growth
of galaxy mass \citep{Vogt97,Boehm03,Milvang-Jensen03,Bamford05}.  One
specific and rather simple test is to determine whether the stellar
mass in galaxies has increased in lock-step with the mass of the dark
matter halo or whether the stellar mass grows within a pre-existing
dark matter halo.  The former is predicted by ``hierarchical'' galaxy
formation models \citep[e.g.\,][]{WhiteFrenk91,Kauffmann94,Cole2k}
since the growth of the stellar disk is regulated by the rate at which
gas is accreted by the halo (the accretion rate is similar for gas and
dark matter).  The latter is characteristic of simple ``classical''
galaxy formation models such as \citet{Eggen62}.  In simplified
outline, the hierarchical model predicts that the correlation between
the terminal rotation velocity and stellar mass should evolve little,
while the classical model predicts that the stellar mass corresponding
to a fixed rotation velocity should decrease with increasing redshift.

This observational test can be made using the galaxies photometry as a
proxy for stellar mass.  It is now generally accepted that the shape,
zero-point and scatter of the TF relation depend on bandpass and that a
smaller dispersion is obtained for near-infrared band-passes (i.e.
rest-frame $I$ rather than rest-frame $B$-band).  The dust-correction
is much smaller in near-infrared photometry and moreover, it is more
sensitive to the underlying, evolved stellar population which best
traces the stellar mass and thus correlates more tightly with the
maximum rotational speed (or total galaxy mass;
\citealt{Verheijen01,Conselice05}) (we note that, for gas-rich systems
the bluer pass-bands are more sensitive to current star-formation and
therefore may also correlate closely to total baryonic mass).

However at $z{\gsim}1$ the small angular size of galaxies means that
obtaining spatially resolved rotation curves is extremely challenging
(e.g.\, a typical local spiral has a scale length of 4\,kpc, which
corresponds to only $0.5''$ at $z{=}1$).  One solution to this problem
is to exploit the magnifying power of foreground galaxy cluster lenses
-- for a typical lens magnification factor of ten, 0.6$''$ corresponds
to an unlensed physical scale of just $\lsim$0.5\,kpc at $z=1$.  Thus
galaxies can be targeted that would otherwise be too small.  Spatially
resolved kinematics can also be achieved on spatial scales far greater
than otherwise possible due to the increased flux sensitivity over
conventional observations.  Moreover, the benefits of gravitational
magnification are complemented by Integral Field Spectroscopy (which
produces a contiguous {\it x, y, velocity} map at each point in the
galaxy) of the target galaxies.  Clean decoupling of the spatial and
spectral information is therefore feasible, thus eliminating problems
arising from mixing of the two in traditional long-slit observations.
It is therefore much easier to identify which galaxies have regular
(bi-symmetric) velocity fields for comparison with local spirals.

In this paper we present a study of six $z{=}1$ gravitationally-lensed
galaxies observed through the cores of four galaxy cluster lenses.
Five of the targets lie behind Abell\,2390, Cl\,2236-04 and
RGB\,1745+398, the sixth (Arc\#289 at $z{=}1.034$ behind A\,2218) was
previously discussed in \citet{Swinbank03}.  All six targets were
observed with the Gemini Multi-Object Spectrograph Integral Field Unit
(GMOS IFU) on Gemini-North.  We concentrate on the galaxy dynamics as
traced by the [O{\sc ii}]$\lambda\lambda$3726.1,3728.8\AA\ emission
line doublet. The IFU data provide a map of the galaxy's velocity field
in sky co-ordinates.  To interpret this field we correct for the
magnification and distortion caused by the lensing potential using
models of the cluster lenses.  These models are constrained by the
positions and redshifts of other spectroscopically confirmed
gravitational arcs in each cluster.  In a few cases, the IFU targets
are themselves multiply imaged and the folding of the velocity field
places additional constraints on the cluster potential.  These lensing
corrections allow us to reconstruct the intrinsic (unlensed) properties
of the galaxies at $z{=}1$, including their geometry.

The velocity field of systems displaying regular (bi-symmetric)
rotational velocity fields i.e.\ resembling rotating disks are reduced
to traditional one-dimensional rotation curves.  The terminal rotation
velocity, indicative of the galaxy mass, is then extracted.  All of the
targets have been imaged in the optical and near-infrared bands.  The
gravitational lens models allow us to correct the photometric
observations for the amplification of the cluster and to establish the
intrinsic (unlensed) rest-frame $B$- and $I$-band luminosities of the
galaxies. We can then compare the TF relations for our target galaxies
with local observations, and hence to test the models for the build-up
of galaxy mass as a function of look-back time.

The structure of this paper is as follows. In \S\ref{sec:obs} we
describe the sample selection, observations and the data reduction.  In
\S\ref{sec:analysis} we analyse the data, including photometry,
gravitational lens modeling and galaxy reconstruction.  In
\S\ref{sec:results} we present the one-dimensional rotation curves and
the TF relation at $z{=}1$.  Finally in In \S\ref{sec:discussion} we
discuss our results on the TF relation at $z{=}1$ and the implications
for galaxy evolution. We also outline the wider applicability of the
gravitational telescope method.  Through-out this paper we use a
cosmology with $H_{0}=72\kms$, $\Omega_{0}=0.3$ and $\Lambda_{0}=0.7$.

\section{Observations and Data Reduction}
\label{sec:obs}

\subsection{Sample Selection} 

%
%
\begin{figure*}
  \centerline{ 
    \psfig{file=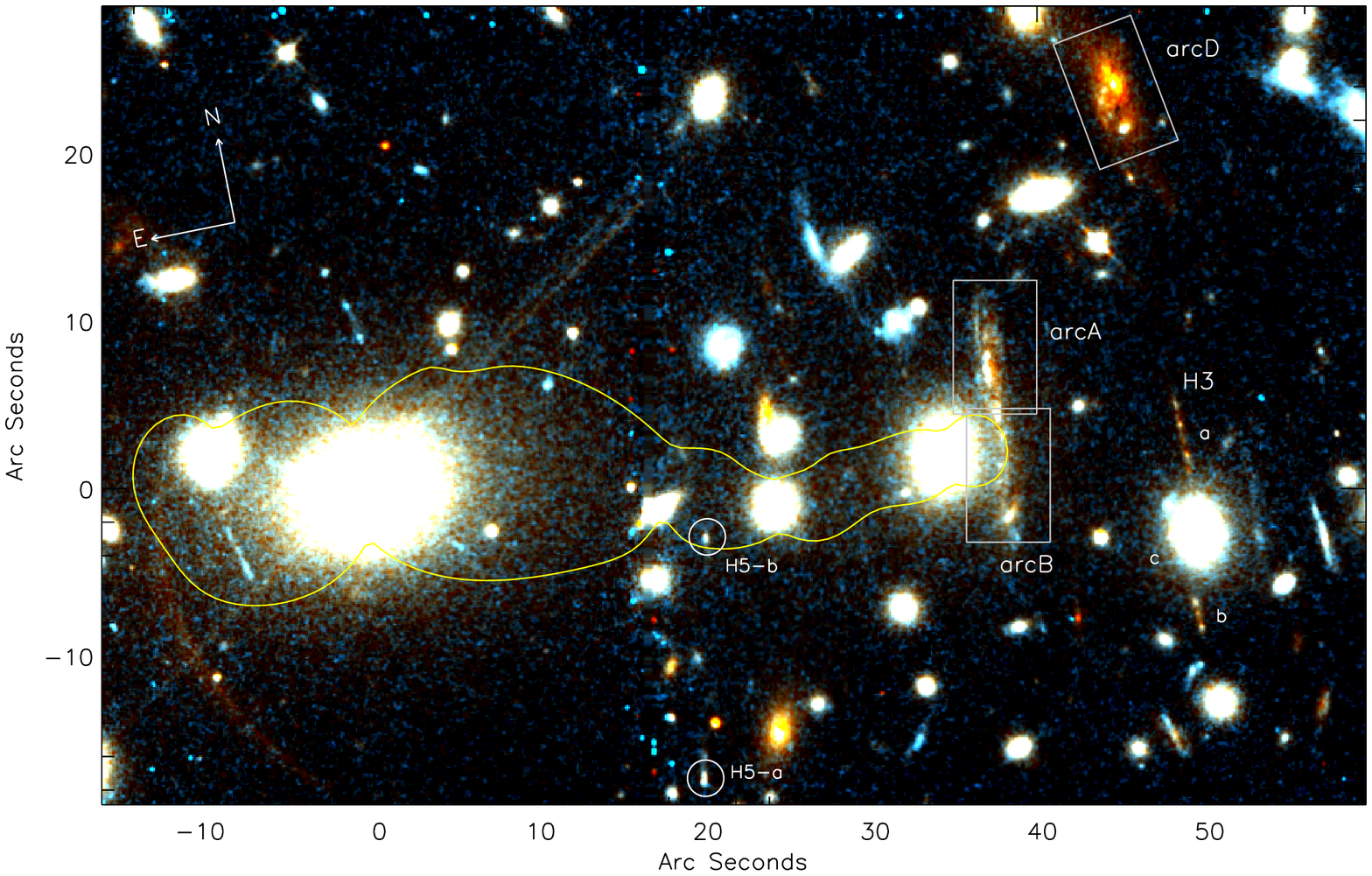,width=6.6in,angle=90}
  }
  \centerline{
    \psfig{file=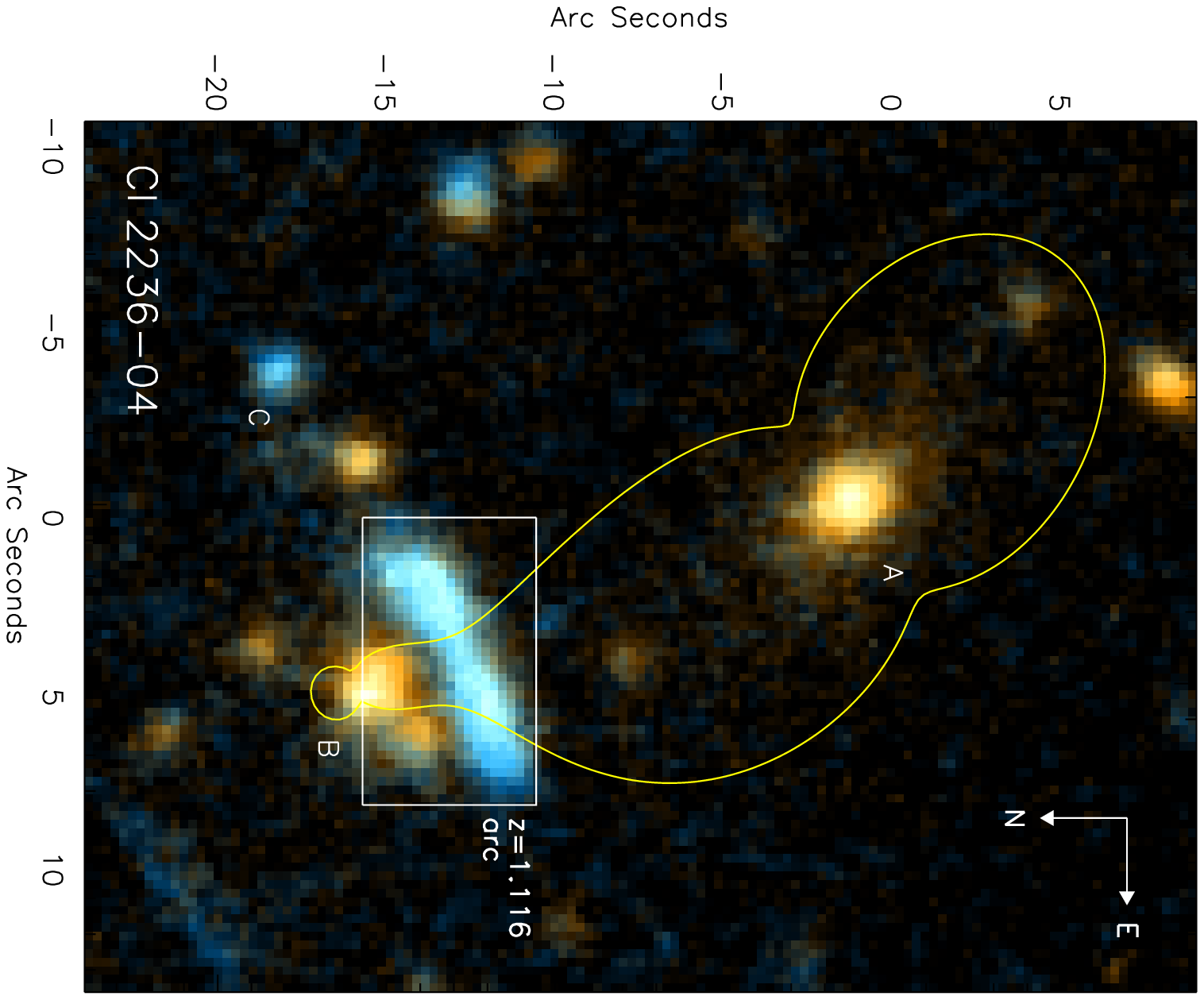,width=3.0in,angle=90}
    \psfig{file=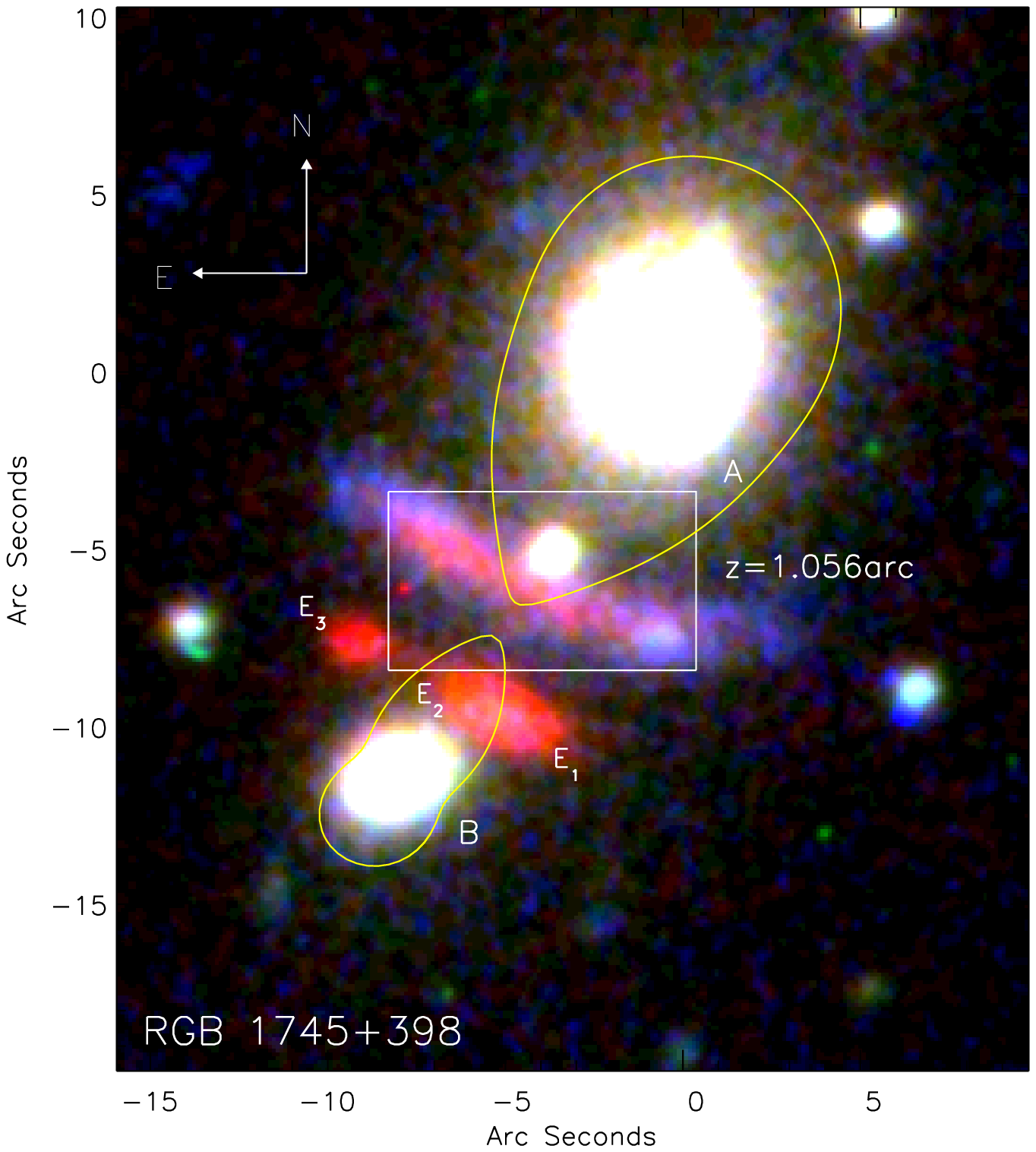,width=3.25in,angle=90}
  }
  \caption{
    Finding charts for the IFU observations.  {\bf Top:} True colour
    {\it HST} $VI$ image of the lensing cluster A\,2390 at $z{=}0.228$
    with the GMOS IFU field of view overlaid on the three $z{\simeq}1$
    arcs.  The yellow curve is the $z{=}0.912$ critical curve
    calculated from the best-fit lens model of this cluster
    (\S\ref{sec:2390}; see also Fig.~\ref{fig:2390}).  We also label
    the $z{=}4.04$ and $z{=}4.05$ multiple-image systems that were used
    to constrain the lens model: H3-a/b/c and H5-a/b respectively.
    {\bf Lower left:} True colour ground-based $BVR$ image of the
    lensing cluster Cl\,2236-04 with the GMOS IFU field of view
    overlaid on the $z{=}1.116$ arc.  The brightest two cluster
    galaxies are labelled $A$ and $B$, and the $z{=}1.334$ source
    discussed by \citet{Kneib94} is labelled $C$.  We also show the
    $z{=}1.116$ critical curve in yellow.  {\bf Lower right:} True
    colour ground-based $BRK$ image of the lensing cluster
    RGB\,1745+398 with the GMOS IFU field of view overlaid on the
    $z{=}1.056$ arc and the $z{=}1.056$ critical curve shown in yellow.
    The multiply-imaged ERO at (-5'',-9'') has a redshift of
    $z{=}1.11{\pm}0.05$ (\S\ref{sec:longslit}) and is labelled $E_{1}$,
    $E_{2}$ and $E_{3}$.  }
  \label{fig:finder}
\end{figure*}

%
%
\begin{table}
\begin{center}
\hspace*{-0.7cm}
{\scriptsize
\smallskip

\hspace*{-0.7cm}\begin{tabular}{lcccccccc}
\hline\hline
\noalign{\smallskip}
Source              & {\it z$_{cl}$} &$\alpha_{\rm J2000}$  &   $\delta_{\rm J2000}$ & {\it z$_{arc}$} & $t_{exp}$  & ref   \\
                    &                &(h m s)               &   ($^{\circ}\ '\ ''$)  &                 & {\it (ks)} &       \\
\hline
\noalign{\smallskip}
A\,2218\,arc\#289   & 0.176         & 16\,35\,55.07        & +66\,11\,51.00         & 1.034           & 5.4        & 1,2,3 \\
RGB\,1745+398arc    & 0.267          & 17\,45\,38.11        & +39\,51\,23.80         & 1.056           & 10.8       & 4     \\
A\,2390arcA         & 0.233          & 21\,53\,34.52        & +17\,42\,02.32         & 0.912           & 14.4       & 6,7   \\
A\,2390arcB         & 0.233          & 21\,53\,34.30        & +17\,41\,56.01         & 1.032           & 12.0       & 5,6   \\
A\,2390arcD         & 0.233          & 21\,53\,34.39        & +17\,42\,21.19         & 0.912           & 10.8       & 7     \\
Cl\,2236-04arc      & 0.560          & 22\,39\,33.00        & -04\,29\,19.83         & 1.116           & 10.8       & 8     \\
\hline
\label{table:arcs}
\end{tabular}
}
\vspace{-0.5cm}
\caption{Redshifts for the cluster lens ($z_{cl}$); arcs ($z_{arc}$)
  and spectroscopic imaging exposure times for the arcs our sample.
  References for arc discovery or previous study: 1: \citealt{Pello92};
  2: \citealt{Ebbels98}; 3: \citealt{Swinbank03}; 4:
  \citealt{Nilsson99}; 5: \citealt{Pello99}; 6: \citealt{Frye98}; 7:
  \citealt{Pello91}; 8: \citealt{Kneib94}
}
\end{center}
\end{table}

The positions and redshifts of the gravitational arcs observed during
this program are listed in Table~1.  In order to avoid possible biases,
the targets were selected to be representative of lensed galaxies in
the distant Universe --- no attempt was made to select galaxies with
relaxed late-type morphology.  However, we did require that arcs were
resolved in both spatial dimensions so that a two dimensional velocity
field could be extracted from the IFU data.  This restricted our
selection to galaxies with moderate magnification.

\subsection{Imaging}
\label{sec:imaging}

Near-infrared imaging of Abell 2390 was obtained the Wide-field
Infrared Camera (WIRC; \citealt{Wilson03}) on the Hale 200$''$
Telescope\footnote{The Hale Telescope at Palomar Observatory is owned
  and operated by the California Institute of Technology.}.  The
$J$-band data were taken on 2004 June 30, totaling 12.8\,ks.  The
$K$-band data were taken over four observing runs between October 2003
and June 2004 yielding a total integration time of 11.5\,ks.  All of
these data were taken when transparency was good, with the $J$-band
conditions likely photometric.  The seeing was typically $\lsim0.85''$.
The data were reduced using standard {\sc iraf}\footnote{IRAF is
  distributed by the National Optical Astronomy Observatory, which is
  operated by the Association of Universities for Research in
  Astronomy, Inc., under a cooperative agreement with the National
  Science Foundation.}  procedures.  Photometric calibration was
achieved using 2MASS\footnote{This paper makes use of data products
  from the Two Micron All Sky Survey (2MASS), which is a joint project
  of the University of Massachusetts and the Infrared Processing and
  Analysis Center/California of Technology, funded by the National
  Aeronautics and Space Administration and the National Science
  Foundation.}  data on stars in the same fields.  We also retrieved
observations of A\,2390 from the \emph{Hubble Space Telescope (HST)}
public archive\footnotemark.  The Wide Field Planetry Camera 2 (WFPC2)
observations through the $I_{814}$- and $V_{555}$-band filters totalled
10.5 and 8.4\,ks respectively -- the data were reduced using the
standard {\sc stsdas} package in {\sc iraf}.  The final reduced data
are shown as a colour image in (Fig.~\ref{fig:finder}).  A\,2390 has
also been observed with the Advanced Camera for Surveys (ACS) as part
of the Guaranteed Time Observations (GO: 9292, PI: Ford).  We
concentrated solely on the $z'$-band ACS data, reducing the single
orbit (2.2\,ks), CR-split, undithered data using {\sc multidrizzle}
with the default parameter set.

\footnotetext{Obtained from the Multimission Archive at the Space
  Telescope Science Institute (MAST).  STScI is operated by the
  Association of Universities for Research in Astronomy, Inc., under
  NASA contract NAS5-26555. Support for MAST for non-HST data is
  provided by the NASA Office of Space Science via grant NAG5-7584 and
  by other grants and contracts.}

Ground-based optical imaging of Cl\,2236-04 and RGB\,1745+398
(Fig.~\ref{fig:finder}) is taken from \citet{Kneib94} and
\citet{Nilsson99} respectively.  These consist of $B,R$ and
$B,V,R,I$-band imaging of these clusters taken with the New Technology
Telescope (NTT) and Nordic Optical Telescope (NOT) respectively.  We
supplement these optical data with near-infrared $J$- and $K$-band
imaging of both clusters on the UK Infra-Red Telescope (UKIRT) between
2004 August 15 and 2004 August 20\footnotemark.  The observations were
made in photometric conditions and $\sim0.7''$ seeing using the UKIRT
Imaging Spectrometer (UIST) imaging camera \citep{RamseyHowat98} which
employs a $1024\times1024$ InSb detector at 0.12$''$\,pixel$^{-1}$ to
give a 2$'$ field of view.  The observations were taken in a standard
nine-point dither pattern and reduced using the relevant {\sc orac-dr}
pipeline \citep{Cavanagh03}.  The total integration times for each band
was 2.4\,ks.  To calibrate our data, we observed UKIRT faint
photometric standards \citep{Howarden01}.  These standards were
observed at similar air masses and using the same instrumental
configuration as the target galaxies.

\footnotetext{The United Kingdom Infrared Telescope is operated by the
  Joint Astronomy Centre on behalf of the U.K. Particle Physics and
  Astronomy Research Council.}

The most striking feature in the new near-infrared imaging is the
discovery of a red triply-imaged galaxy 4$''$ to the South-East of the
$z{=}1.056$ arc in RGB\,1745$+$398 (Fig.~\ref{fig:finder}).  This arc
is barely detected in the optical imaging, but is very bright in the
near-infrared ($J=18.27\pm0.06$ and $K=16.32\pm0.04$) and an $R-K$
colour of $6.6\pm0.2$.  The obvious triple image configuration, along
with a spectroscopic redshift of this arc would provide a new
constraint on the lens model of this cluster and therefore a more
precise reconstruction of the intrinsic properties of the blue arc
targeted by the IFU observations (\S\ref{sec:ifu_obs}).  We describe
ground-based optical and near-infrared spectroscopy of the red arc in
\S\ref{sec:longslit}.

\subsection{Integral Field Spectroscopy}
\label{sec:ifu_obs}

\begin{figure*}
  \centerline{\psfig{figure=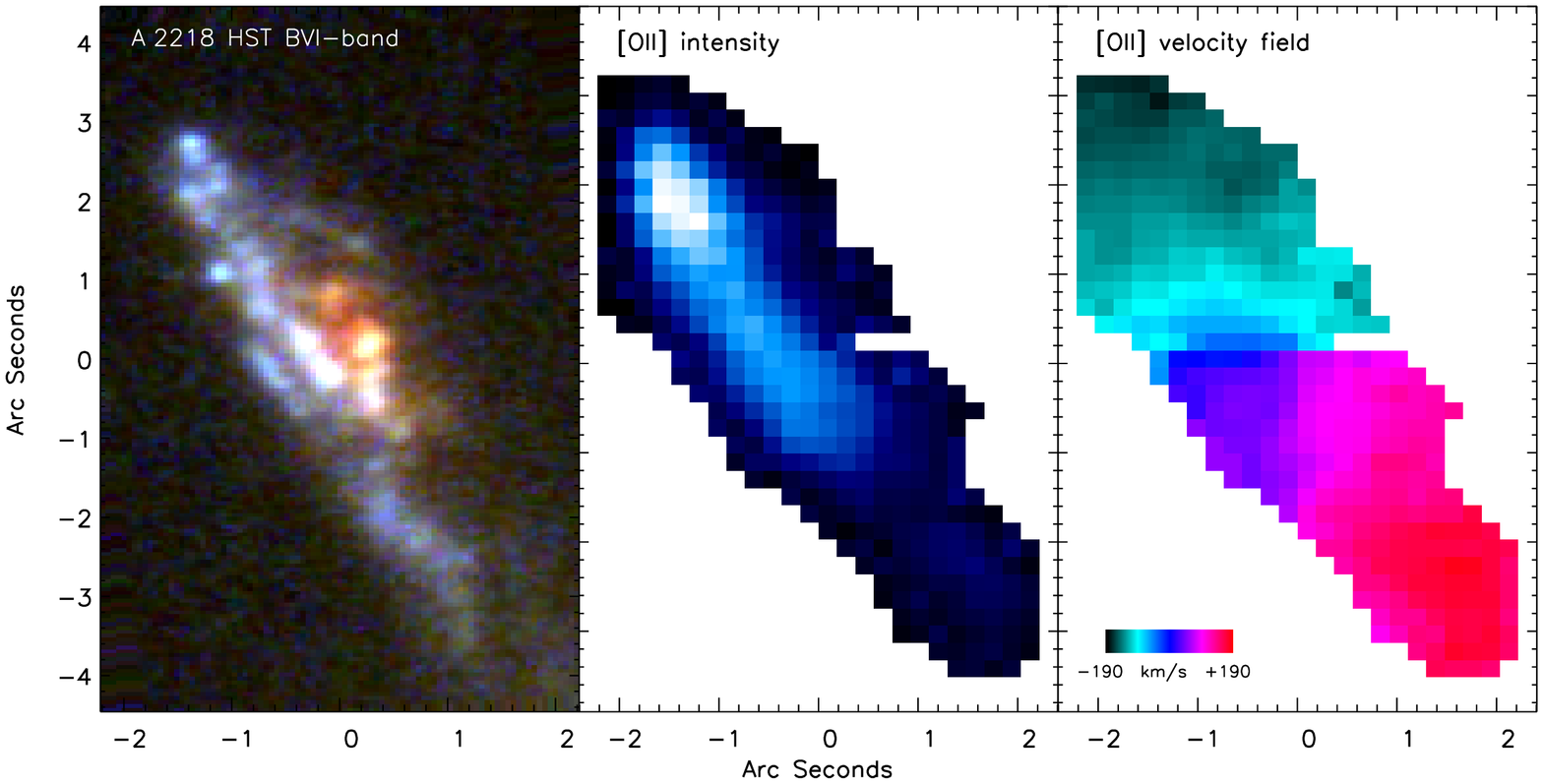,width=4.5in,angle=90}}
  \centerline{\psfig{figure=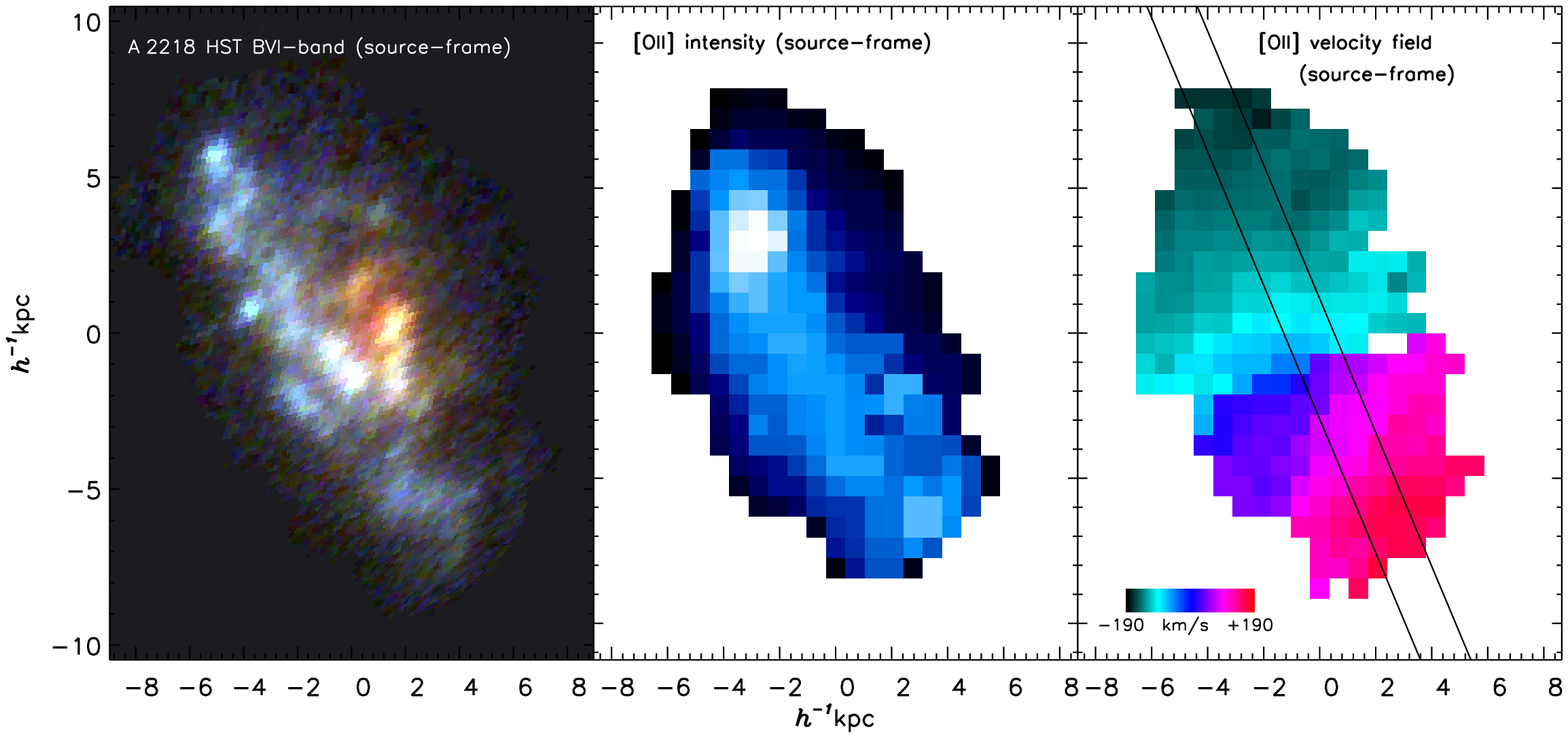,width=4.5in,angle=90}}
  \caption{ {\bf Top:} \emph{HST}/WFPC2 and GMOS IFU observations of
    A\,2218~arc\#289. {\it Left:} \emph{HST}/WFPC2
    $B_{450}V_{606}I_{814}$ image of A2218 arc\#289. {\it Middle:} The
    [O{\sc ii}]\,$\lambda$3727 emission map of the arc measured from
    our IFU observations. The distribution of [O{\sc ii}] emission
    agrees well with the rest-UV flux seen in the left panel. {\it
      Right:} The velocity field of the galaxy derived from the [O{\sc
        ii}] emission. The scale is marked in arcseconds and North is
    up and East is left.  {\bf Bottom:} The reconstructed galaxy after
    correcting for lens magnification -- see \S\ref{sec:2218} for
    details. {\it Left:} the reconstructed image of the galaxy. {\it
      Middle:} The reconstructed [O{\sc ii}] emission line map. {\it
      Right:} The reconstructed velocity map.  The red and blue
    regions represent redshift and blueshift respectively, and the
    solid lines show the asymptotic major axis cross section from
    which the one-dimensional rotation curve was extracted
    (\S\ref{sec:onedRC}). The scale shows the size of the galaxy after
    correction for lensing.}
  \label{fig:2218}
\end{figure*}

\begin{figure*}
  \centerline{
    \psfig{figure=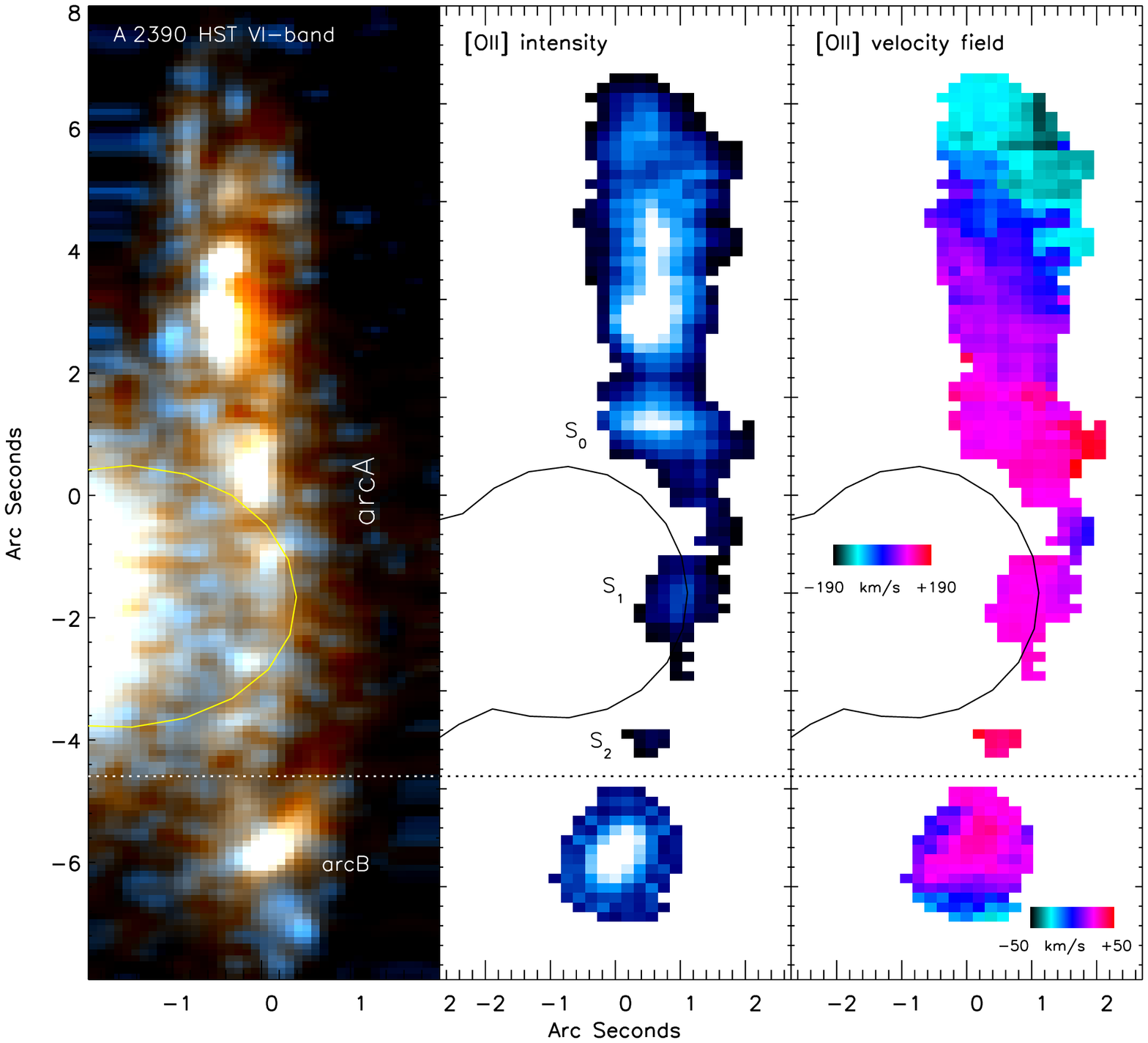,width=3.5in,angle=90}
    \psfig{figure=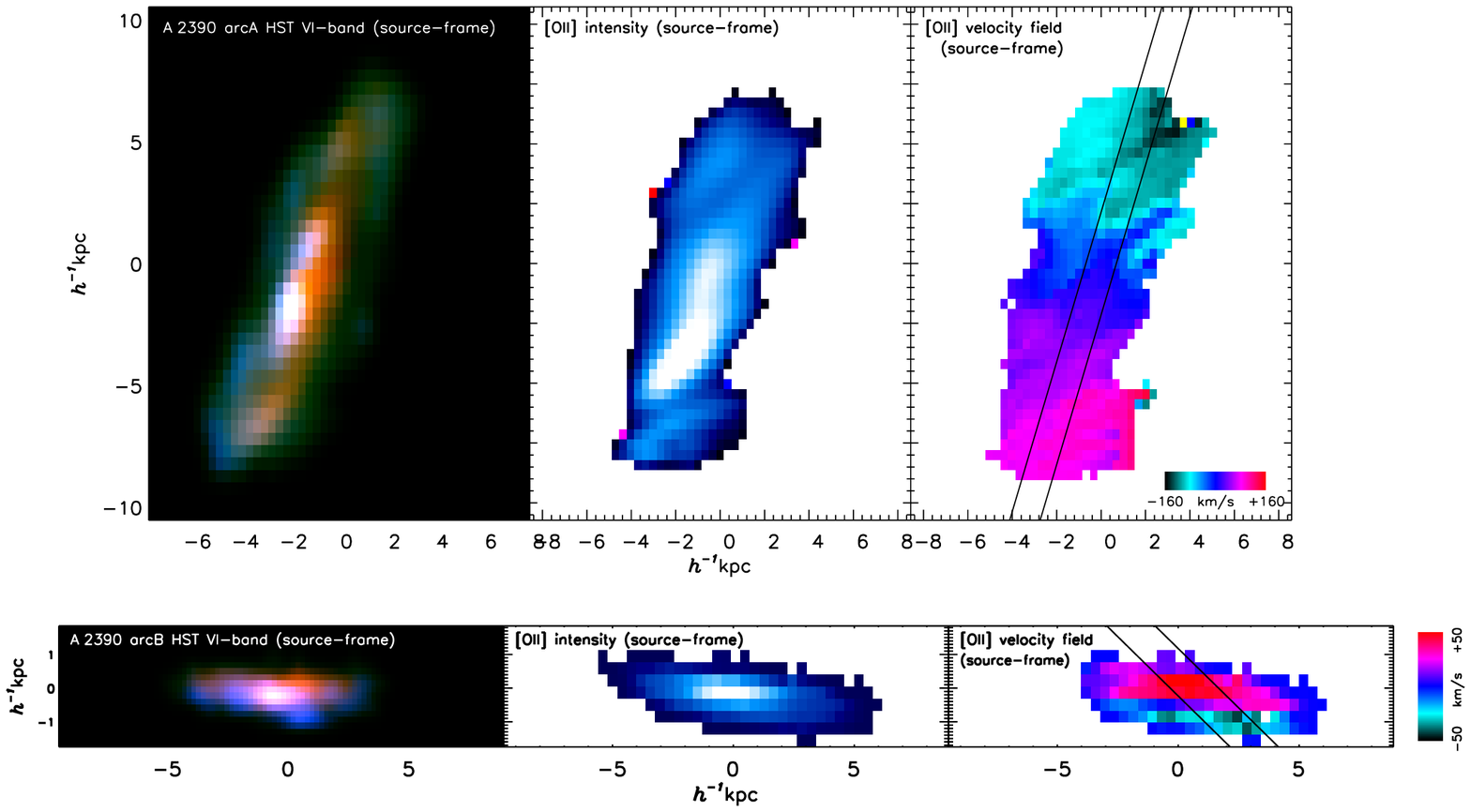,width=4in,angle=0}
  }
  \centerline{\psfig{figure=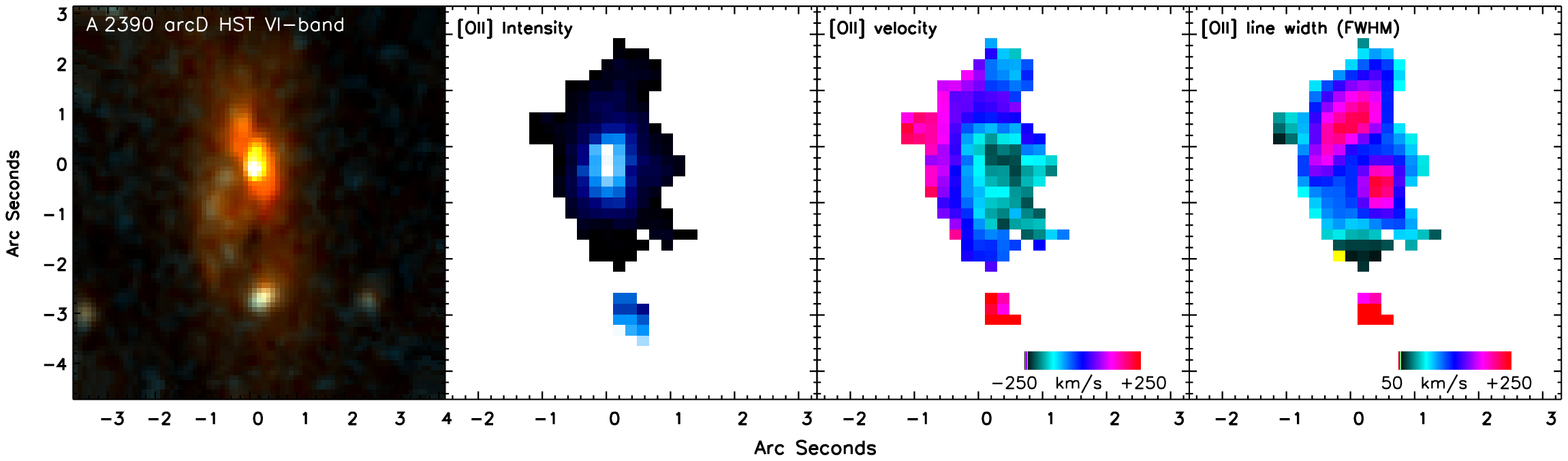,width=6in,angle=90}}
  \centerline{\psfig{figure=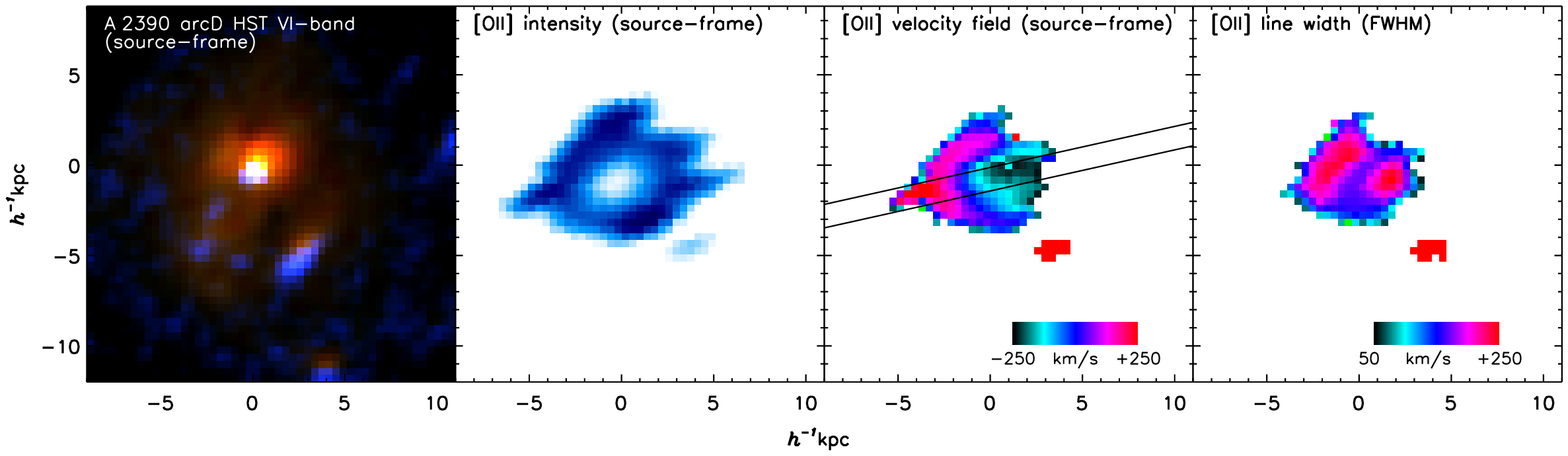,width=6in,angle=90}}
  \caption{
    {\bf Top left:} \emph{HST} and IFU observations of A\,2390arcA and
    arcB. We show the $V_{606}I_{814}$- band image of the two arcs
    (left), the [O{\sc ii}] emission map (middle) and the velocity
    field of both galaxies (right). The distribution of [O{\sc ii}]
    emission in both galaxies agrees well with the rest-UV flux seen in
    the \emph{HST} imaging. In all three panels the $z{=}0.912$
    critical curve is shown in yellow/black.  {\bf Top right:} The
    reconstructed galaxies after correcting for lens magnification --
    see \S\ref{sec:2390} for details. The upper and lower three panels
    show the reconstructed image, [O{\sc ii}] emission line map and the
    velocity field of arcA and arcB respectively.  In the velocity maps
    the red and blue regions represent redshift and blueshift
    respectively, and the solid lines show the asymptotic major axis
    cross section from which the one-dimensional rotation curve was
    extracted.  {\bf Middle:} \emph{HST} and IFU observations of
    A\,2390arcD. {\it Far left:} $V_{606}I_{814}$-band color picture
    showing the complex morphology of this red galaxy. {\it Left
      center:} The [O{\sc ii}] emission line map. The distribution of
    [O{\sc ii}] emission line flux is well matched to the broad-band
    imaging, and confirms that the bright knot (located $3''$ to the
    south) is associated with this galaxy. {\it Right center:} The
    velocity field derived from the [O{\sc ii}] emission. {\it Far
      right:} The [O{\sc ii}] emission line width (FWHM) map which
    shows a large variation in the distribution of line widths (both in
    the spatial and spectral domain).  {\bf Bottom:} The reconstructed
    galaxy after correction for lens magnification. The galaxy has a
    complex morphology, with at least two components. The peak-to-peak
    velocity gradient of the galaxy is orthogonal to the major axis
    seen in the broad-band imaging. Furthermore, the distribution of
    the [O{\sc ii}] line widths show two (off-center) peaks
    approximately along the same direction as the maximum velocity
    gradient. The dense knot or companion located 7 kpc to the south
    appears redshifted by $300{\pm}80\kms$ and has a line width of
    ${\sim}300\kms$ FWHM.  }
  \label{fig:2390}
\end{figure*}

\begin{figure*}
  \centerline{\psfig{figure=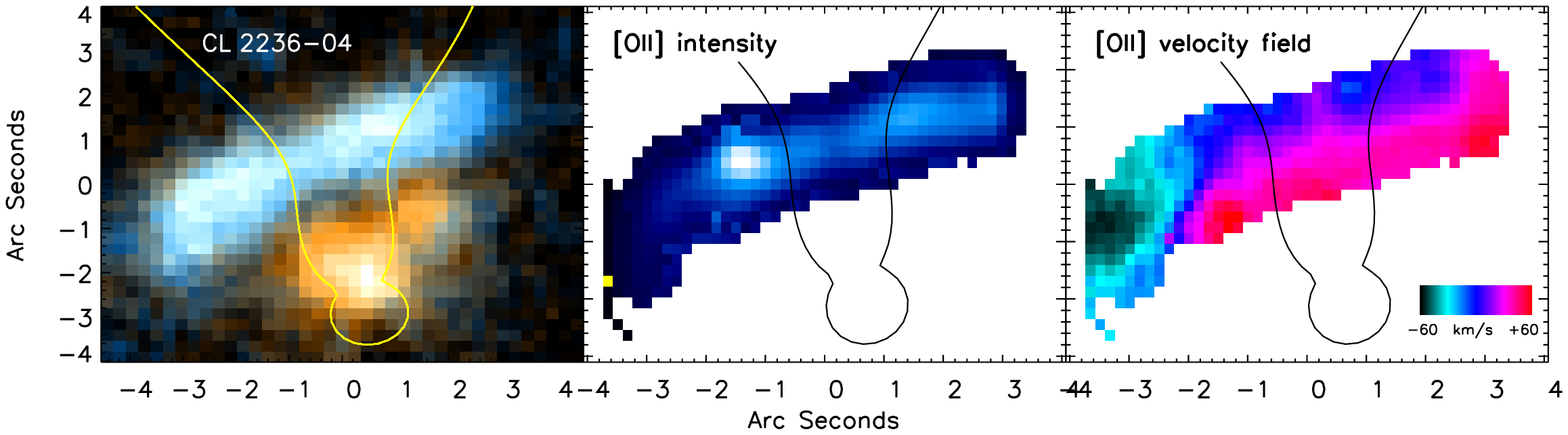,width=5in,angle=90}}
  \centerline{\psfig{figure=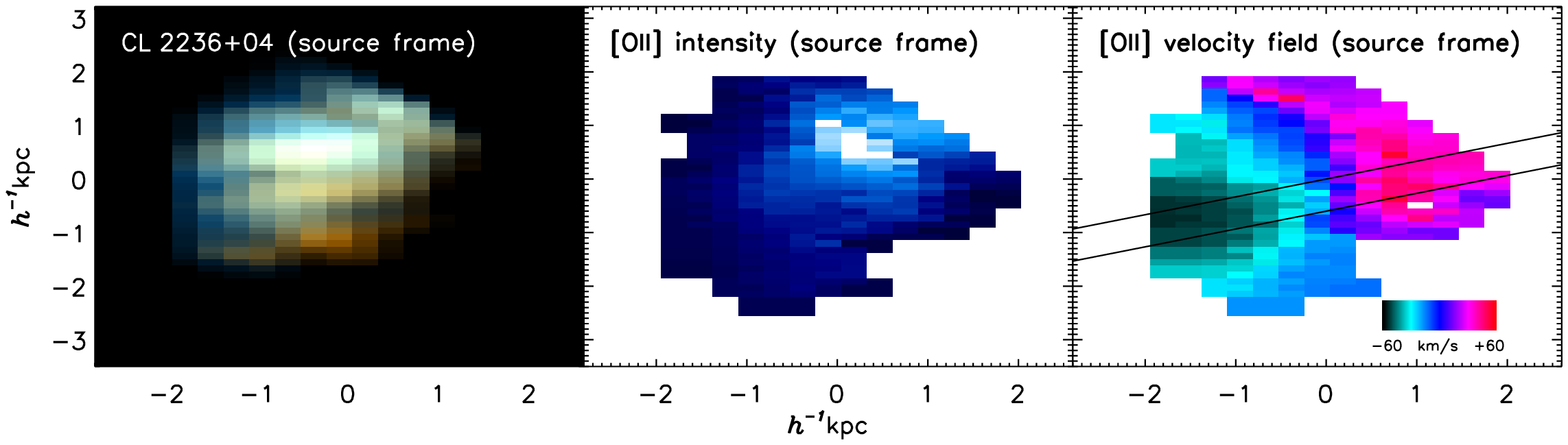,width=5in,angle=90}}
  \caption{
    {\bf Top:} Ground-based imaging and IFU observations of the
    $z{=}1.116$ arc in Cl 2236-04. {\it Left:} $BVR$-band colour
    picture. {\it Middle:} [O{\sc ii}] emission line map . {\it Right:}
    The velocity field derived from the [O{\sc ii}] emission.  In all
    three panels the $z{=}1.116$ critical curve of the best-fit lens
    model is shown in yellow/black. {\bf Bottom:} The reconstructed
    galaxy after correction for lens magnification. {\it Left:} the
    reconstructed $BVR$-band colour picture. {\it Middle:} The
    reconstructed [O{\sc ii}] emission line map. {\it Right:} The
    reconstructed velocity map. The red and blue regions represent
    redshift and blueshift respectively, and the solid lines show the
    asymptotic major axis cross section from which the one-dimensional
    rotation curve was extracted.  }
  \label{fig:2236}
\end{figure*}

\begin{figure*}
  \centerline{\psfig{figure=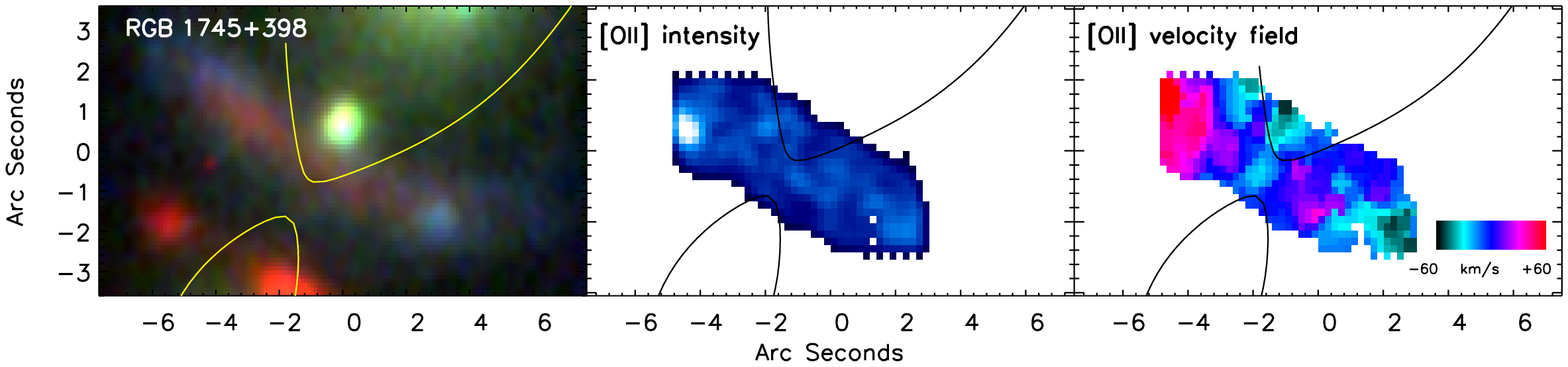,width=6in,angle=90}}
  \centerline{\psfig{figure=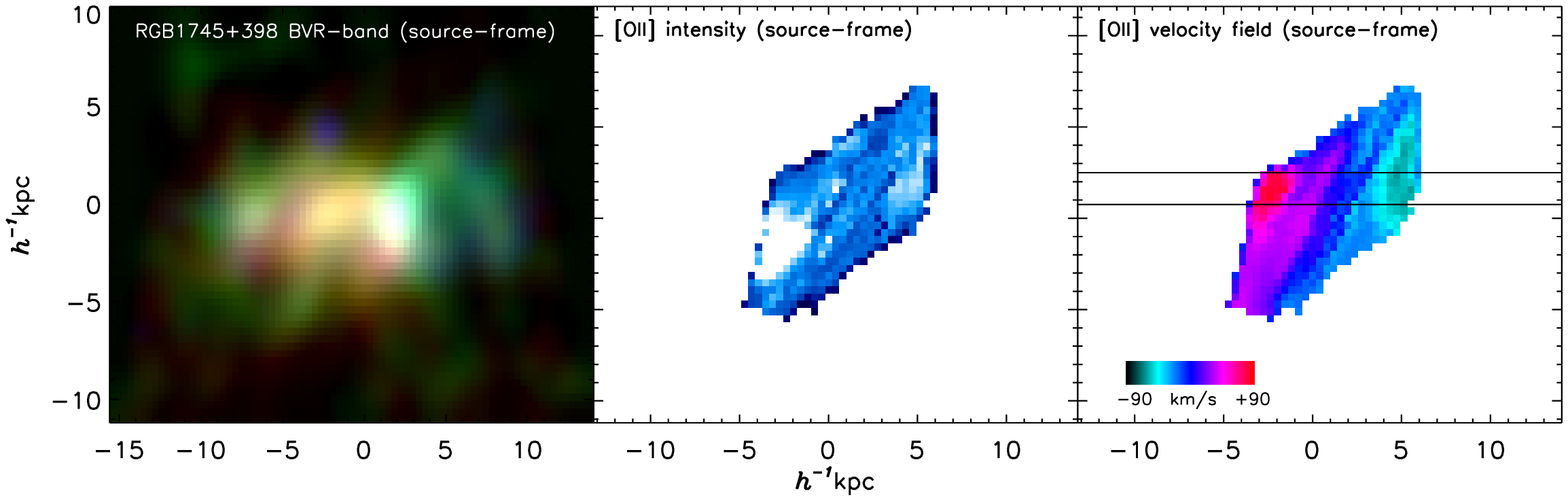,width=5in,angle=90}}
  \caption{
{\bf Top:} Ground-based imaging and IFU observations of the
$z{=}1.056$ arc in RGB1745+398. {\it Left:} Ground-based colour
$BRK$-band colour picture. {\it Middle:} [O{\sc ii}] emission line
map. {\it Right:} The velocity field derived from the [O{\sc ii}]
emission. In all three panels the $z{=}1.056$ critical curve from the
best-fit lens model is shown in yellow/black. {\bf Bottom:} The
reconstructed galaxy after removal of lens magnification.  {\it Left:}
Reconstructed $BRK$-band colour picture. {\it Middle:} The
reconstructed [O{\sc ii}] emission line map: note that the morphology
seen in this image is largely determined by the IFU field of
view. {\it Right:} The reconstructed velocity map. The red and blue regions
represent redshift and blueshift respectively, and the solid lines
show the asymptotic major axis cross section from which the
one-dimensional rotation curve was extracted.  }
  \label{fig:1745}
\end{figure*}

Traditionally, observations are made with a two-dimensional detector,
sufficient for imaging programs.  However, for spectroscopic
measurements, one spatial dimension of the sky is usually lost in order
to allow dispersion of the image across one axis of the detector.
Spectrograph's placing a long slit over the object lose spatial
information orthogonal to the slit width.  In order to faithfully
investigate the properties of galaxies which are spatially extended, we
require spectroscopic information over the whole of a two-dimensional
field.  An instrument simultaneously producing both two-dimensional
imaging data and a spectrum for each point in the image provides
integral field spectroscopy.

Integral field spectroscopy thus allows spectra to be simultaneously
obtained for a number of contiguous areas across a two-dimensional
field.  The spatially resolved spectroscopy allows the variation with
positions of object spectra to be measured, and this information can be
used, for example, to determine the object's internal kinematics, or
the star formation rate within an object, providing insight into its
process of evolution.  This technique is therefore crucial in order to
investigate the internal properties of distant galaxies.

Our sample were observed in queue mode with the Gemini Multi-Object
Spectrograph (GMOS) on Gemini-North\footnotemark\ between 2003 June 8
and 2003 August 14 in photometric conditions and $\lsim$0.7$''$ seeing.
Details of the observations and source redshifts are listed in Table~1.
The IFU uses a lensed fiber system to reformat the a contiguous $7.9''
\times 5.3''$ field (comprising an array of 40$\times$26 hexagonal
fibres each of 0.2$''$ diameter) into two long slits
\citep{AllingtonSmith02}.  All observations were made using an $I$-band
filter in conjunction with the R400 grating which results in two tiers
of spectra covering the maximum field of view.  The spectral resolution
of this configuration is $\lambda/\Delta\lambda =2000$.  Each
observation was split into 2.4\,ks sub-exposures and dithered by one
IFU lenslet to account for bad pixels.  The data was reduced in {\sc
  iraf} using the GMOS-IFU data reduction pipeline which extracts,
flat-fields and wavelength calibrates the data.  To subtract the sky
emission lines the GMOS IFU employs a second ($5''\times3''$) IFU
separated by one arcminute on the sky.  We used {\sc idl} to identify
and extract sky-fibres adjacent to object fibres on the spectrograph
and used these to achieve the sky-subtraction. We also improved the
flattening of the data by using continuum regions either side of the
emission lines.  This is achieved by identifying wavelength regions
free from sky emission lines near the redshifted [O{\sc ii}] line and
averaging the flux in these regions over 200\AA.  For each fiber, this
average was divided into the fiber spectrum to improve the
fiber-to-fiber response map.  The output pixel scale is 1.3\AA\ pixel
and the instrumental profile has a FWHM of 3.4\AA\ (measured from the
widths of the skylines).  This corresponds to $\sigma=58\kms$ in the
galaxy rest-frame, and the set-up adequately resolves the [O{\sc
  ii}]$\lambda\lambda$3726.1,3728.8\AA\ emission-line doublet.  In all
following sections we have deconvolved the instrumental resolution from
the line width measurements where these could be reliably determined.

\footnotetext{Programme ID: GN-2003A-Q-3. The GMOS observations are
  based on observations obtained at the Gemini Observatory, which is
  operated by the Association of Universities for Research in
  Astronomy, Inc., under a cooperative agreement with the NSF on behalf
  of the Gemini partnership: the National Science Foundation (United
  States), the Particle Physics and Astronomy Research Council (United
  Kingdom), the National Research Council (Canada), CONICYT (Chile),
  the Australian Research Council (Australia), CNPq (Brazil) and
  CONICET (Argentina).}

For one of the targets, (A\,2390arcA), we required two pointing's to
cover the large spatial extent ($\sim12''\times4''$) of the galaxy.
During the second pointing the much smaller arc ($\sim2''\times2''$ at
$z=$1.033; A\,2390arcB) was also covered.  In order to align and mosaic
these two datacubes we constructed (wavelength-collapsed) white-light
[O{\sc ii}] emission line maps of the two arcs.  This was achieved by
identifying the central wavelength of the [O{\sc ii}] emission in each
galaxy and then collapsing the data-cubes between -300 and +300$\kms$.
These were then compared and aligned using the {\it HST} $I$-band image
as a reference to produce the final mosaiced datacube.

We also note that the redshift of the arc in RGB\,1745+398 ($z$=1.056)
places the [O{\sc ii}] emission at 7660\AA; close to the Fraunhofer
A-band.  To correct for this absorption, we extracted the spectrum of
the foreground cluster galaxy (which is a strong continuum source at
this wavelength; see \S\ref{sec:1745}) and use this to model and
correct for the telluric absorption.

After constructing the datacubes, we proceed to fit the [O{\sc ii}]
emission line doublet in each pixel using a $\chi^2$ minimisation
procedure, taking into account the greater noise at the position of the
sky lines.  The spectra were averaged over 3$\times$3 spatial pixels
($0.6''\times0.6''$), increasing this region to 4$\times$4 pixels
($0.8''\times0.8''$) if the signal was too low to give a sufficiently
high $\chi^2$ improvement over a fit without the line.  In regions
where this averaging process still failed to give an adequate $\chi^2$,
no fit was made.  Using a continuum fit, we required a minimum $\chi^2$
of 25 (S/N of 5) to detect the line and when this criterion is met, we
fit the [O{\sc ii}] emission line doublet with a double Gaussian
profile of fixed separation, allowing the normalization and central
wavelength to vary. If the line is detected at greater than 7$\sigma$
significance, we also allow the line width to vary. The two lines of
the doublet are assumed to have the same intensity and width.  Whilst
it is possible for the two emission lines of the [O{\sc ii}] doublet to
have different intensities (due to the density and temperature of the
gas), at this spectral resolution and typical signal-to-noise we view
it as more reliable to fix the ratio of the intensities to unity (we
note that in the collapsed spectrum from each galaxy there is no
noticable difference between in the fluxes between the two peaks).  To
calculate the error in the velocity we perturb the wavelength of the
best fit profile and allow the signal to drop by a $\chi^2$ of 9.  This
corresponds to a formal 3$\sigma$ error.

\subsection{Long Slit Spectroscopy}
\label{sec:longslit}

%
%
\begin{figure}
  \centerline{\psfig{file=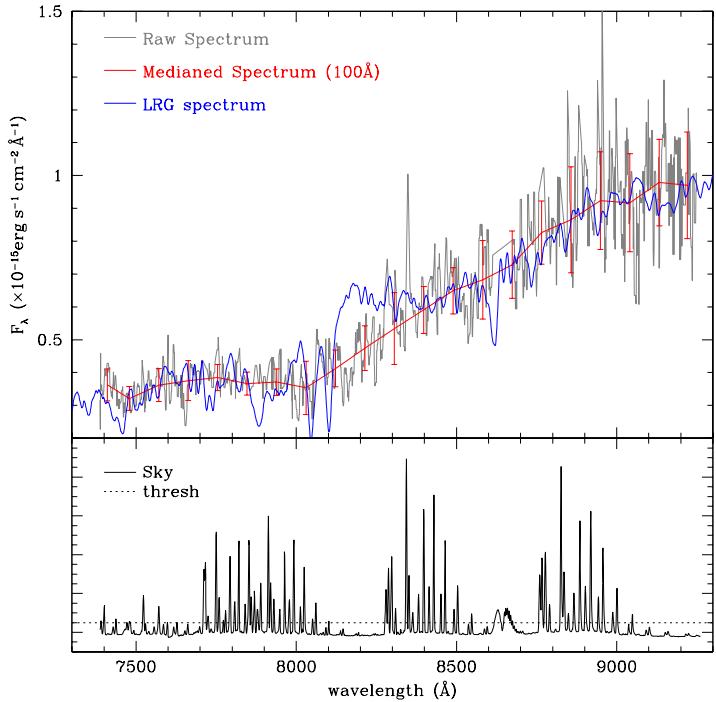,width=4.0in,angle=0}}
  \caption{LRIS-R spectra of the multiply imaged ERO in RGB\,1745+398.
    The spectrum is produced by rejecting pixels dominated by the
    strong night sky emission i.e. those regions above the horizontal
    line in the lower spectrum which shows the night sky spectrum
    (offset and scaled for clarity).  The smoothed spectrum is produced
    by median filtering the raw spectrum with a boxcar width of 100\AA\ 
    and the error bars represent 1$\sigma$ errors in each boxcar
    window.  We identify the discontinuity in the spectral shape at
    $\lambda$8440\AA\ as the 4000\AA\ break which yields a redshift of
    $z{=}1.11\pm0.05$.  The blue line shows an Luminous Red Galaxy
    (LRG) galaxy template from the Sloan Digital Sky Survey redshifted
    to $z{=}1.11$.}
  \label{fig:rgb1745_ero_spec}
\end{figure}

In the course of assembling the new and archival imaging data required
for this article we discovered a new triply-imaged red arc in
RGB\,1745$+$398 (Fig.~\ref{fig:finder}).  We estimate the redshift of
this arc using {\sc hyper-z} \citep{Bolzonella00} and the photometry
given in \S\ref{sec:imaging}, obtaining $z=1.1\pm0.2$.  We attempted to
measure a spectroscopic redshift for this arc on 2005 February 22.  We
used the Cooled Grating Spectrograph 4 (CGS4) and UIST, both on UKIRT
to explore $1.0{-}1.2{\mu\rm m}$ and $1.4{-}2.4{\mu\rm m}$ respectively
-- the total integration times for both of these observations was 1
hour.  We detected continuum emission in these data, but no obvious
line emission.  We therefore also searched for spectral features at
shorter wavelengths using the red arm of the Low Resolution Imaging
Spectrograph (LRIS; \citealt{Oke95}) in long slit mode on the Keck I
10-m telescope\footnotemark on 2005 March 05.  The 831/8200 grating was
oriented at $36^\circ$ to sample the wavelength range
${\sim}0.75{-}0.9{\mu{\rm m}}$.  A total integration time of 3.3\,ks
was accumulated from three separate exposures with a $1''$ wide slit.
Flux calibration was achieved via observations of the
spectrophotometric standard G138-31 \citep{Oke90}.  The data were
reduced using standard {\sc iraf} tasks -- the final reduced
1-dimensional spectrum is shown in Fig.~\ref{fig:rgb1745_ero_spec}.
Again, no strong emission or absorption features were found, however
there is a continuum break at 8440\AA\ which we interpret as the
4000\AA\ break redshifted to $z=1.11\pm0.05$.  The uncertainty on the
redshift is estimated by comparing the data with the mean spectrum of a
luminous red galaxy from the Sloan Digital Sky Survey;
(Fig.~\ref{fig:rgb1745_ero_spec}).

\footnotetext{Obtained at the W.M. Keck Observatory, which is operated
  as a scientific partnership among the California Institute of
  Technology, the University of California and the National Aeronautics
  and Space Administration. The Observatory was made possible by the
  generous financial support of the W.M. Keck Foundation.}

\section{Data Analysis and Modelling}
\label{sec:analysis}

\subsection{Photometry}

%
%
\begin{table*}
{\tiny
\begin{center}
{\centerline{\sc Table~2: Arc Photometry}}
\begin{tabular}{lcccccc}
\hline\hline
                  &                  \multicolumn{6}{c}{Photometry}                                                     \\
                  &   $B$          &   $V$          &   $R$          &      $I$       &   $J$          &   $K$          \\
                  &                &                &                &                &                &                \\
\hline
A\,2218\,arc\#289 & $21.66\pm0.04$ & $20.86\pm0.10$ & $20.53\pm0.04$ & $19.79\pm0.05$ & $18.63\pm0.06$ & $17.32\pm0.06$ \\
RGB1745+398arc    & $21.90\pm0.12$ & $21.26\pm0.10$ & $20.90\pm0.07$ & $19.6\pm0.3$   & $18.29\pm0.09$ & $16.72\pm0.07$ \\
A2390arcA         &    ...         & $21.64\pm0.06$ &    ...         & $19.91\pm0.04$ & $18.64\pm0.10$ & $17.42\pm0.08$ \\
A2390arcB         &    ...         & $23.34\pm0.07$ &    ...         & $21.65\pm0.05$ & $20.10\pm0.10$ & $19.41\pm0.06$ \\
A2390arcD         & $24.55\pm0.17$ & $21.62\pm0.04$ &    ...         & $19.53\pm0.03$ & $17.31\pm0.03$ & $15.64\pm0.02$ \\
Cl2236-04arc      & $20.50\pm0.11$ &   ...          & $20.05\pm0.11$ &   ...          & $19.47\pm0.14$ & $18.09\pm0.12$ \\
\hline
\label{table:photom}
\end{tabular}
\end{center}
}
{\addtolength{\baselineskip}{-3pt}\vspace{-0.3cm}
  {\scriptsize
 \noindent{\bf Table~2.} Optical and near-infrared photometry for the arcs in our sample.
}}
\end{table*}

\setcounter{table}{2}
%
%
\begin{table*}
{\tiny
\begin{center}
{\centerline{\sc Table~3: Source-Frame Properties of the Arcs}}
\begin{tabular}{lcccccccc}
\hline\hline
                  & Amplification          & Inclination & v$_{rot}$   &    M$_{B}$      &     M$_{I}$     & \multicolumn{3}{c}{Disk Scale Lengths}   \\
                  & $(\mu)$                & {\it (i)}   &   $\kms$    &                 &                 &   $h_{V}$ & $h_{I}$ & $h_{[OII]}$        \\
                  &                        &             &             &                 &                 &           & $(kpc)$ &                    \\
\hline
A\,2218\,arc\#289 & $4.92^{+0.20}_{-0.15}$ & $64\pm6$    & 206$\pm$17  & -21.10$\pm$0.20 & -22.34$\pm$0.21 &    2.4    &   2.1   & 2.2                \\
RGB1745+398arc    & $14.8^{+4.4}_{-6.6}$   & $60\pm7$    & 103$\pm$25  & -19.55$\pm$0.60 & -20.22$\pm$0.55 &     -     &    -    &  -                 \\
A2390arcA         & $12.6^{+0.6}_{-0.8}$   & $69\pm4$    & 170$\pm$28  & -20.09$\pm$0.24 & -21.37$\pm$0.25 &    2.0    &   2.2   & 2.6                \\
A2390arcB         & $18^{+1}_{-1}$         & ...         &   ...       & -18.24$\pm$0.30 & -18.70$\pm$0.20 &    0.8    &   0.8   & 1.1                \\
A2390arcD         & $6.7^{+0.4}_{-0.2}$    & ...         &   ...       & -21.77$\pm$0.20 & -23.27$\pm$0.15 &     -     &    -    &  -                 \\
Cl2236-04arc      & $11.2^{+2.8}_{-2.4}$   & $45\pm10$   & 98$\pm$30   & -19.79$\pm$0.22 & -20.51$\pm$0.23 &    1.6    &   1.3   & 1.9                \\
\hline
\label{table:sourceproperties}
\end{tabular}
\end{center}
}
\caption{Notes: $\mu$ denotes luminosity weighted magnification.  We
estimate the 1--$\sigma$ uncertainty in $\mu$ by perturbing the
parameters of the best fit lens model such that $\Delta\chi^{2}=1$ and
for each model we recompute the magnification of the arc.  The error
corresponds to the largest variation in magnification in the source
plane photometry.  The inclination of the galaxies are measured in the
source-frame and the associated error-bars are calculated by
determining the inclination in each pass-band and for each of the lens
models described above.  The rotation velocities (v$_{rot}$) are
corrected for inclination effects.  The disk scale lengths are given
for the reconstructed $I$ and $V$-band images as well as the
distribution of [O{\sc ii}] flux.
}
\end{table*}

From our optical/near-infrared imaging, we constrain the spectral
energy distribution (SED) of each galaxy.  Since the arcs usually lie
with a few arcs-seconds of nearby bright cluster galaxies, we calculate
the magnitude of the arcs in various pass-bands by masking the arc and
interpolating the light from the nearby cluster members.  We then use
the {\sc iraf imsurfit} package with the sky estimated from a 2$^{\rm
  nd}$ order polynomial surface fit.  This is subtracted from the
regions around each galaxy to account for the halo light from the
nearby cluster galaxies.  We then use {\sc sextractor} \citep{Bertin96}
to estimate the residual background within the frame and extract the
arc photometry using {\sc sextractor}.  We use the observed colours (at
the known redshift) to infer the ratio of current to past star
formation rate and use this to find the best fit SED.  At $z\sim1$ the
rest frame $I$-band luminosity is approximately equivalent to the
observed $H$-band and so is calculated from interpolating $J$ and $K$
photometry.  We co-add the $J,K$ images to obtain an aperture which is
used to extract the $J+K$ magnitudes and interpolate for the relevant
SED type.  We apply the same technique for the rest-frame $B$-band
magnitude, (at $z\sim1$, $R$ and $I$-bands are the closest match to
rest-frame $B$-band) and estimate its uncertainty by computing the
magnitude for a variety of SED types which are consistent with the
observed optical colours.

\subsection{Gravitational Lens Modelling and Galaxy Reconstruction}
\label{grav_tel}

In order to investigate the source-plane properties of the sources we
must first correct for the distortion and magnification of the galaxy
image by the cluster lens.  Although this does not affect the sign or
amplitude of the velocities, we must determine accurately the
gravitational magnification of each galaxy so that we can recover its
intrinsic geometry and hence estimate the inclination angle.  In this
section we describe briefly the construction of the cluster lens models
and their application to derive the intrinsic properties of the
galaxies at $z{=}1$.  The lens modelling techniques are those described
in detail by \citet{Smith05}, originally developed by
\citet{Kneib93PhD}, and further refined by \citet{Kneib96} and
\citet{Smith02Th}.  The primary constraints on the lens models are the
positions and redshifts of spectroscopically confirmed gravitational
arcs in each cluster.  The IFU data (\S\ref{sec:ifu_obs}) provide
additional constraints because the components of the observed arcs that
are identified as being multiple-images of the same region of the
respective lensed galaxies must, within the uncertainties, have the
same velocity and [O{\sc ii}] line strength, in addition to the usual
broad-band flux and colour constraints.

\subsubsection{A\,2218}
\label{sec:2218}

Arc~\#289 in A2218 was discussed in detail by \citet{Swinbank03}. Here
we summarize the key details and use this arc to explain the methods
applied to the other arcs.

A\,2218 is one of the best constrained strong lensing clusters,
including three multiple-image systems that have been spectroscopically
confirmed via their line emission.  This cluster has been modeled on
numerous occasions: Kneib et al.\ (1993; 1996; 2004); Ellis et al.\ 
(2001); Smith et al.\ (2005).  We use the most recent of these to
calculate the luminosity weighted magnification ($\mu$).  This was
achieved by ray-tracing between the image- and source-planes to build
up a map across the arc of how the observed flux relates to the
intrinsic (unlensed) flux.  The statistical error on $\mu$ was derived
from the family of lens models which adequately reproduce the
multiply-imaged arcs.  For each model we then recompute the
magnification of each arc.  The error corresponds to the largest
variation in magnification that we found via this method.  The
luminosity weighted amplification for this arc is
${\mu}{=}4.92^{+0.20}_{-0.15}$, which translates to a boost in
magnitude of $\Delta m=1.7\pm0.1$.

The mapping between image- and source-plane co-ordinates described
above was also used to reconstruct the intrinsic morphology of Arc\#289
from the flux map of the arc after lensing correction, (see Fig.~3).
The reconstruction reveals that Arc\#289 is a blue disk-galaxy with
much internal structure, resembling a late-type galaxy.  This is
especially prominent in the $B$-band which samples the rest-frame UV
and is therefore dominated by the star-forming {H\sc ii} regions
(Fig.~3).  To measure the geometry of the disk we fitted ellipses to an
isophote in the reconstructed flux maps using the {\sc idl gauss2dfit}
routine (and assume an intrinsically circular disk).  The ellipticity
is then $e = 1 - b / a$ (where $a$ and $b$ are the major and minor axis
of the ellipse) and the inclination, $i$, is $\cos i = b / a$.  The
average axis ratios of the ellipses from the various passband are
translated into inclinations and the error bars on $i$ are computed by
applying the same procedure to each passband and each of the lens
models described above.  The inclination of the galaxy is found to be
$64\pm6^{\circ}$.

We use the intrinsic velocity field (constructed in a manner similar to
that applied to the broad-band imaging above) to infer the rotational
velocity of the galaxy's disk.  The terminal rotation velocity is
$186\pm16\kms$ (i.e. half the asymptotic velocity shift across the
galaxy).  Using the inclination angle derived above, we therefore
calculate a corrected rotation velocity of $v_{rot}{=}(206\pm17)\kms$.

\subsubsection{A\,2390}
\label{sec:2390}

A cluster lens model of A\,2390 was originally developed by
\citet{Pello99} who spectroscopically identified two z$\sim$4 multiply
imaged galaxies behind the cluster.  The first (labelled H3 in
Fig.~\ref{fig:finder}) has a redshift of $z=4.04$ and all three images
lie within 10$''$ of the cluster galaxy to the West.  The second lies
at $z=4.05$ (labelled H5-a and H5-b in Fig~\ref{fig:finder}).  However,
the third image of this galaxy was not identified by \citet{Pello99}
because it lies outside of the field of view of the \emph{HST}/WFPC2
observations available at that time.  We have used the larger field of
view of the new ACS $z'$-band data (\S\ref{sec:imaging}) to identify
the third counter-image of H5 at $\alpha$=21:53:36.64,
$\delta$=+17:42:11.8 (J2000).  These spectroscopically confirmed
multiple-image systems are used to constrain the lens model of this
cluster, including the generation of a family of acceptable models as
described in \S\ref{sec:2218}.  The $z=0.912$ critical curve of the
best-fit model is shown in Fig.~\ref{fig:finder} -- this curve defines
where multiply-imaged galaxies at $z$=0.912 will appear in the observed
(image) plane.  In a simple axi-symmetric system the gravitational
magnification is infinite along critical curves.  This is not the case
in the perturbed gravitational potential of a galaxy cluster,
nevertheless magnifications of ${\sim}10{-}20{\times}$ are routine
(Table~2).

\smallskip
\noindent{\bf A2390\,arcA} -- This arc (the so-called ``straight arc'' --
\citealt{Pello99}) is ${\sim}10''$ long, has a redshift of $z=0.912$
and lies approximately 40$''$ to the West of brightest cluster galaxy
(BCG).  Based solely on the broad-band imaging, the arc \emph{may} be
multiply-imaged at its southern end, indeed the $z{=}0.912$ critical
curve of the best-fit lens model is immediately adjacent to the
southern portion of this arc (Figs.~\ref{fig:finder}~\&~3).  However
the acceptable models include several that predict this arc to be
strongly-sheared at its southern end rather than multiply-imaged.  The
velocity field derived from the GMOS observations supports the
multiple-image interpretation: the southern end of the galaxy ($S_{0}$)
has an observed velocity (with respect to the center of the galaxy) of
$\sim150\kms$ (Fig.~3).  The relative velocity of components $S_{1}$
and $S_{2}$ also agree within $1{\sigma}$ with the velocity of $S_0$.
We conservatively include both multiple-image and strong-shear
interpretations when estimating uncertainties below.

We use the same methods as described in \S\ref{sec:2218} to reconstruct
the intrinsic properties of this arc, obtaining the lensing corrected
morphology, [O{\sc ii}] map and velocity map shown in
Fig.~\ref{fig:2390}.  This galaxy therefore appears to have disk-like
kinematics, to be magnified by ${\mu}{=}12.6^{+0.6}_{-0.8}$,
(${\Delta}m=2.75\pm0.07$), and have an inclination angle of
$i{=}(69{\pm}4)^{\circ}$.  The terminal velocity (uncorrected for
inclination effects) is $175\pm20\kms$, which translates into a
corrected rotation velocity of $v_{rot}{=}(187{\pm}17)\kms$.

\smallskip\noindent{\bf A\,2390arcB} -- The GMOS observations of
A\,2390arcA also allowed us to study the [O{\sc ii}] emission from a
slightly higher redshift ($z{=}1.033$) arc, located $2''$ to the South
of A\,2390arcA (Fig.~\ref{fig:finder}~\&~\ref{fig:2390}).  This galaxy
has a compact observed morphology (${\rm FWHM}{\lsim}1.5''$).  We
calculate a luminosity-weighted magnification of $18.0\pm1.0$,
(${\Delta}m=3.1\pm0.1$) which yields a lensing corrected half-light
radius of just $2.5\,{\rm kpc}$.  There is no significant velocity
gradient along the major axis of the reconstructed galaxy (Fig.~3),
however we extract the maximum velocity shear from the data to derive
an estimated a possible peak-to-peak velocity of $60{\pm}15\kms$.
Moreover, the [O{\sc ii}] emission doublet is well resolved with a FWHM
of ${\lsim}60\kms$ (deconvolved for instrumental resolution).  These
estimates place A\,2390arcB comfortably within the scatter of H{\sc ii}
regions in the velocity-width versus half light radius from lower
redshift ($z{\sim}0.19{-}0.35$) narrow-emission line galaxies from
\citet{Guzman96}.

\smallskip\noindent{\bf A2390\,arcD} -- This $z{=}0.912$ arc lies
$19''$ away from A\,2390arcA, their similar redshift suggesting that
A\,2390 may be lensing a group of galaxies.  The \emph{HST}
observations reveal an irregular morphology and the colours are
consistent with either an evolved stellar population or a
dusty-reddened star-burst.  The strong and spatially resolved [O{\sc
  ii}] emission revealed by our GMOS data appear to favor the latter
interpretation, and reveal that the galaxy is not relaxed
(Fig.~\ref{fig:2390}).  The mean velocity of the line emission suggests
possible rotation about the {\it major} axis of the galaxy with
peak-to-peak velocity of $\sim500\kms$.  This is orthogonal to the
direction of rotation suggested by the galaxy's morphology.
Furthermore, the line profiles in the central region appear clearly
broadened, having FWHM${\sim}300\kms$.  This suggests either (i) a
merger between two galaxies, or (ii) shock heated, outflowing material
from the central regions of the galaxy (powered by AGN activity).  Our
observations also confirm that the (unresolved) bright knot (located
$3''$ to the south) is part of the same system, but show that it is
offset by ${\simeq}480{\pm}60\kms$ in velocity and has an [O{\sc ii}]
line width of $300{\pm}100\kms$.  The magnification of this source is
${\mu}{=}6.7^{+0.4}_{-0.2}$ (${\Delta}m=2.0\pm0.1$).

\subsubsection{Cl\,2236-04}\label{sec:2236}

The blue $z{=}1.116$ arc in Cl\,2236$-$04 \citep{Melnick93} lies
between two bright elliptical galaxies (labelled $A$ and $B$ in
Fig.~\ref{fig:finder}). The arc is almost straight with a length of
$8''$, and was shown by \citet{Kneib94} to not be part of a
multiple-image system including the blue source labelled $C$ in
Fig.~\ref{fig:finder}, which is at $z{=}1.334$. \citeauthor{Kneib94}
also identified a velocity gradient along the $z{=}1.116$ arc using
longslit spectroscopic data; but the lack of full two-dimensional data
lead them to suggest that the arc comprises a pair of closely
interacting galaxies. The lack of both high resolution HST imaging and
integral field spectroscopy of the arc made it difficult for them to
test the galaxy-galaxy merger hypothesis. With the full two-dimensional
coverage our GMOS data are therefore a powerful tool in this regard
because they allow us to decouple cleanly the spatial and spectral
information that are mixed in long slit observations.
Fig.~\ref{fig:2236} shows the [O{\sc ii}] emission line intensity and
velocity field of the arc. The velocity field across the arc is seen to
be smooth and regular, with the eastern (receding) portion stretched by
the gravitational potential of the cluster.  There is no obvious sign
of a galaxy-galaxy merger in these data. The simplest interpretation is
therefore that the arc comprises a single disk-like galaxy, the
receding portion of which is either multiply-imaged or strongly-sheared
by the gravitational potential of the foreground cluster -- i.e.\ a
similar interpretation to that of arcA in A\,2390.

We construct a suite of nine models that are able to reproduce the
observed arc morphology and velocity field. All of the models contain
mass components for galaxies $A$ (the BCG) and $B$, plus a mass
component for the central cluster dark matter halo centered on the BCG.
Seven of the models have bi-modal dark matter distribution and contain
a dark matter halo centered on galaxy $B$. The models with a unimodal
dark matter distribution (e.g.\ ${\sigma}_{DM,A}{=}930\kms$,
${\sigma}_{DM,B}{=}0\kms$) shear the galaxy at $z{=}1.116$ sufficiently
to reproduce a symmetric source-plane morphology with respect to the
position where the observed velocity field changes sign. The more
extreme of the bi-modal models (e.g.\ ${\sigma}_{DM,A}{=}750\kms$,
${\sigma}_{DM,B}{=}500\kms$) multiply-image the receding part of the
galaxy, but do not have quite enough mass to multiply-image the
approaching part of the galaxy. The best fit lens model has
${\sigma}_{DM,A}{=}875\kms$, ${\sigma}_{DM,B}{=}300\kms$.  The
$z{=}1.116$ critical curve of the best-fit model is shown in
Figs.~\ref{fig:finder}~\&~\ref{fig:2236}.  All nine models are used
when estimating the uncertainties in the galaxy reconstruction.

Following the methods described in \S\ref{sec:2218}, we use the suite
of models to obtain a mean luminosity weighted magnification of
${\mu}{=}11.2^{+2.8}_{-2.4}$ (${\Delta}m{=}2.6{\pm}0.2$) and a lensing
corrected inclination of $i{=}(45{\pm}10)^\circ$. The observed velocity
profile shows a terminal velocity of $68{\pm}15\kms$ across the galaxy
which translates into an inclination corrected rotation velocity of
$v_{rot}{=}98^{+30}_{-26})\kms$.

\subsubsection{RGB\,1745+398}\label{sec:1745}

RGB\,1745$+$398 is qualitatively similar to Cl\,2236$-$04 as a
gravitational lens in that a straight blue arc lies between the BCG and
the second brightest cluster galaxy (Fig.~\ref{fig:finder}). The arc
was first discovered by \citet{Nilsson99} and spectroscopically
confirmed to be at $z{=}1.056$.  Our new GMOS IFU observations confirm
that the bright source $1''$ to the North-West of the arc is a
foreground cluster galaxy and is not part of the arc.

We construct a suite of lens models to explore the balance of mass
between the dark matter halos centered on galaxies $A$ and $B$ in a
manner analogous to that described in \S\ref{sec:2236}. The model
parameters are constrained by the triply-imaged red arc discussed in
\S\ref{sec:longslit}, and the requirement to reproduce the observed
morphology and velocity field of the blue arc. The best-fit model has
${\sigma}_{DM,A}{=}560\kms$, ${\sigma}_{DM,B}{=}400\kms$, although
formally a very wide range of parameter space is allowed; the extrema
models have ${\sigma}_{DM,A}{=}660\kms$, ${\sigma}_{DM,B}{=}90\kms$ and
${\sigma}_{DM,A}{=}0\kms$, ${\sigma}_{DM,B}{=}630\kms$ respectively. We
note that ${\sigma}_{DM,A}{=}0\kms$ is somewhat unphysical, however
rather than invoke a prior we conservatively use the full range of
models allowed by the data alone. The critical curve of the best-fit
model is shown in Fig.~\ref{fig:finder}.

The velocity map of this galaxy exhibits a strong velocity
gradient. The IFU only covered the central regions of the galaxy (as
can be seen in Fig.~\ref{fig:finder}) with some low surface brightness
emission lying outside the field of view. However, the velocity field
(and rotation curve) is smooth and characteristic of a rotating system
and gives a good indication of the terminal rotation velocity, which
we determine to be $90{\pm}20\kms$. Using the family of models allowed
by the data we derive a luminosity weighted magnification of
${\mu}{=}14.8^{+4.4}_{-6.6}$ (${\Delta}m{=}2.9^{+0.3}_{-0.6}$), a
lensing corrected inclination angle of $i{=}(69{\pm}4)^\circ$ and an
inclination corrected rotation velocity of $v_{rot}{=}103^{+25}_{-23}\kms$.

\section{Results}\label{sec:results}

\subsection{One-Dimensional Rotation Curves}
\label{sec:onedRC}

%
%
\begin{figure*}
  \centerline{\psfig{file=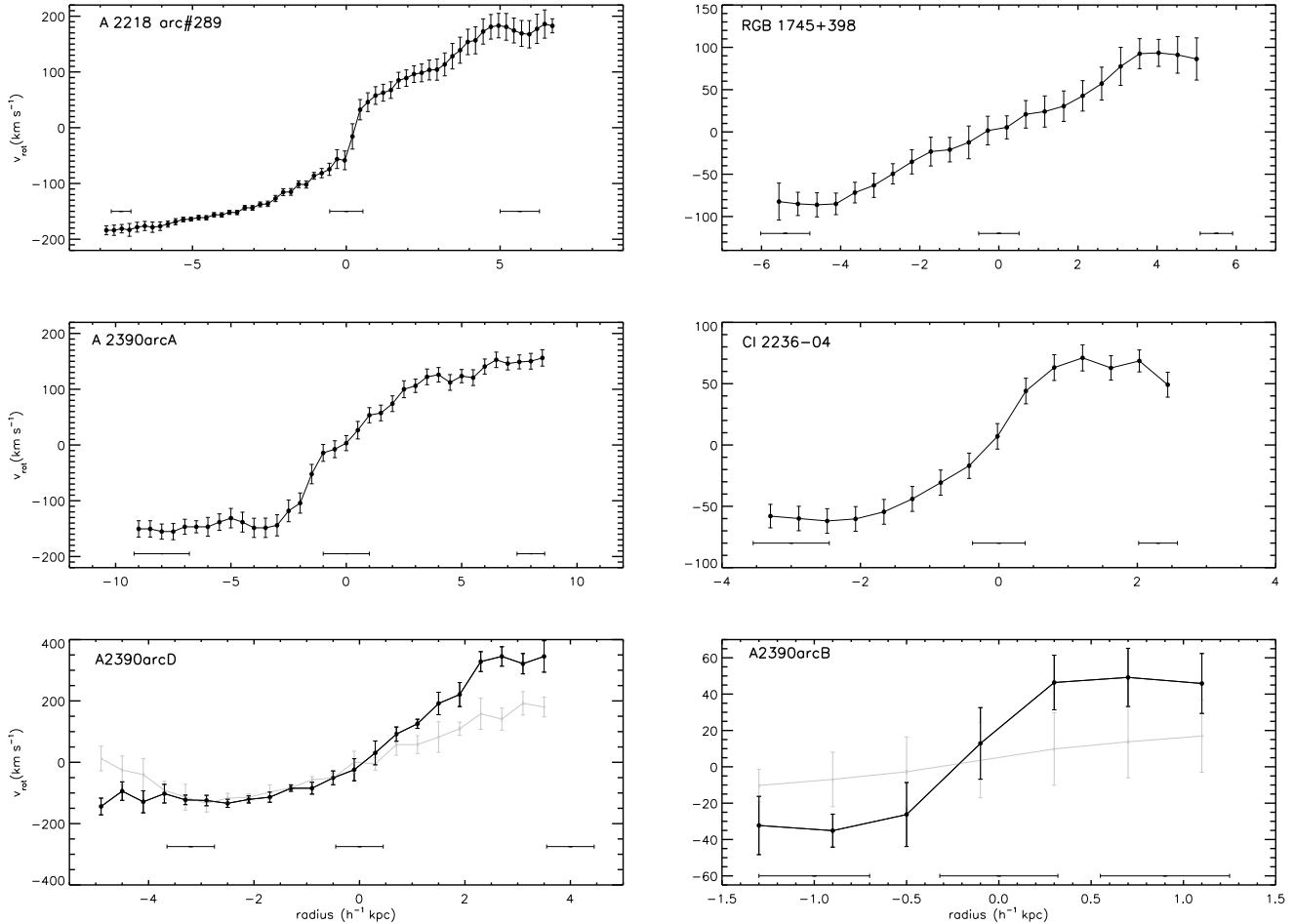,width=7.0in,angle=0}}
\caption{One-dimensional rotation curves from
  the galaxies in our sample.  These are extracted from the
  two-dimensional source plane major axis cross section (and are not
  corrected for inclination effects).  We also show the one-dimensional
  velocity fields for the two galaxies which do not resemble relaxed
  systems: A\,2390arcD and A\,2390arcB.  We extract two one-dimensional
  velocity fields from these two galaxies: the first by collapsing the
  two-dimensional velocity field along the strongest velocity gradient
  (black) as well as the one-dimensional velocity gradient made by
  collapsing the two-dimensional velocity field along the major axis
  seen in the [O{\sc ii}] emission line map (grey).  The solid bars
  represent 0.7$''$ seeing (transformed to the source plane) of each
  galaxy.}
\label{fig:rot_curves}
\end{figure*}

Having reconstructed the source frame morphologies and velocity fields
of the lensed galaxies, we can extract the one-dimensional rotation
curves in order to determine their asymptotic rotation speeds as well
as more general issues such as how the rotation curve shapes compare
with those in the local Universe.  Our sample consists of six galaxies,
the five presented in this paper, plus A\,2218\,arc\#289 presented in
\citet{Swinbank03}.  In four of the lensed galaxies in our sample,
A\,2390arcA, A\,2218\,arc\#289, Cl\,2236-04arc and RGB\,1745+398arc we
identify and extract the major axis velocity profiles to measure the
asymptotic rotation speed. This direction coincides with the maximum
velocity gradient as we would expect in a rotating system.  The other
two galaxies, A\,2390arcB and A\,2390arcD have velocity fields which
are not aligned with the major axis. Although the velocity field is not
consistent with a simple rotating disk, we are still able to place limits
on their masses using the velocity offsets between merging components
(A\,2390arcD) and the limits on velocity gradients/line widths
(A\,2390arcB).

In Fig.~\ref{fig:rot_curves} we show the one dimensional rotation
curves of the galaxies in our sample. These are extracted by sampling
the velocity field with a slit approximately 1.5\,kpc wide along the
major axis cross section, as shown in the reconstructed velocity fields
of the galaxies (Figs. 2-6).  The zero-point in the velocity is defined
using the center of the galaxy in the reconstructed source plane image.
The error bars for the velocities are derived from the formal $3\sigma$
uncertainty in the velocity arising from Gaussian profile fits to the
[O{\sc ii}] emission in each averaged pixel of the datacube.  We note
that in the one-dimensional rotation curves in
Fig.~\ref{fig:rot_curves} alternate points show independent data.  We
also note that the data have not been folded about the zero velocity so
that the degree of symmetry can be assessed in these plots.  Horizontal
bars show the effect of seeing, transformed into the source-plane. This
illustrates how much the velocity field is smoothed by the seeing, and
how the smearing varies along the arc.

%
%
\begin{figure*}
  \centerline{\psfig{file=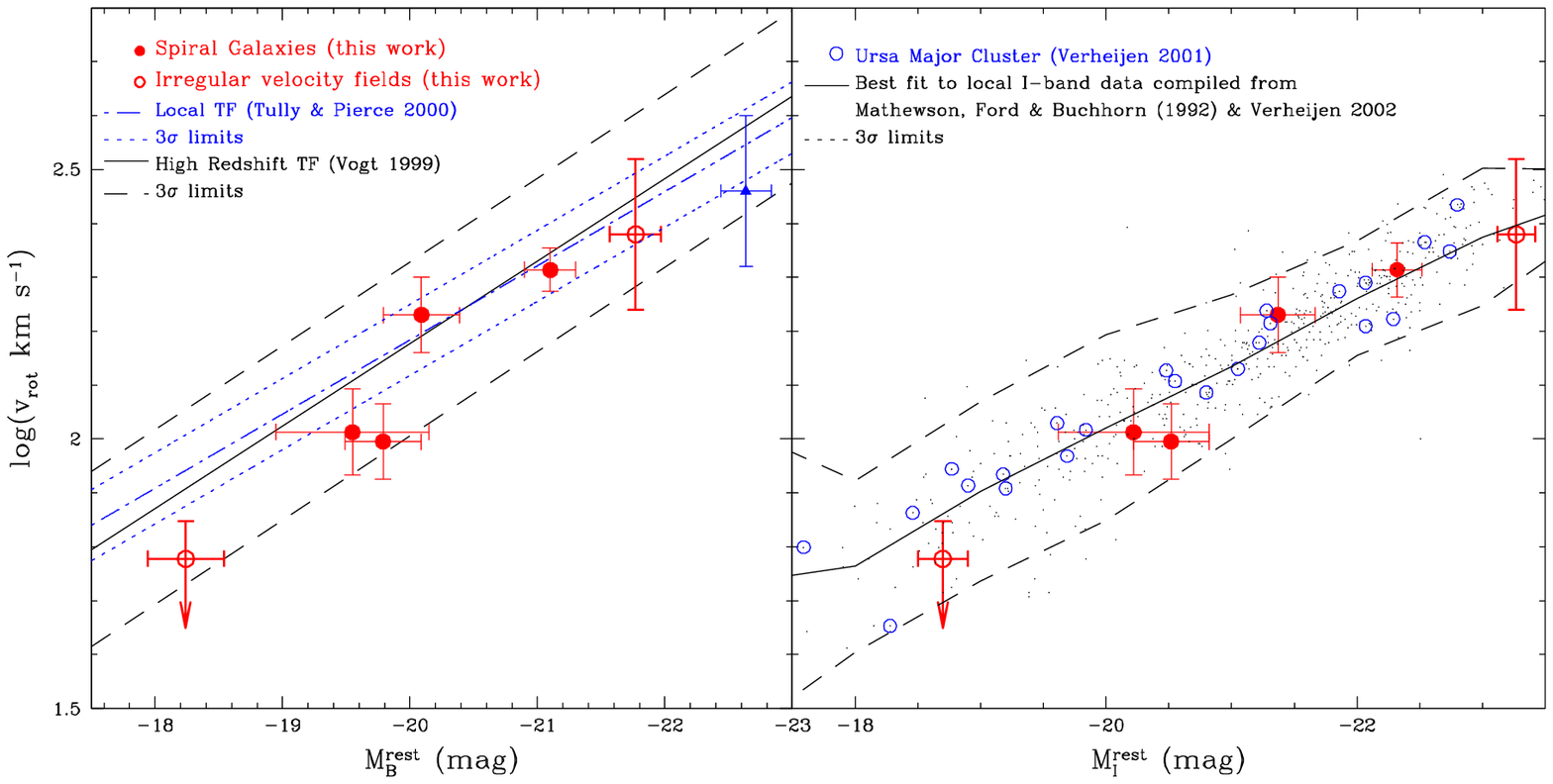,width=7.0in,angle=0}}
\caption{{\bf Left:} The arcs from our survey on the Tully-Fisher
  relation in rest frame $B$-band compared to high redshift sample from
  Vogt et al.\ (1999).  For comparison we show the low redshift local
  fit from Tully \& Pierce (2000).  The solid points show the galaxies
  which have regular (disk-like) kinematics, but we also indicate the
  two galaxies for which the dynamics are not well defines as open
  circles.  The solid triangle shows field galaxies (L451) at $z=1.34$
  from van Dokkum \& Stanford (2001).  {\bf Right:} The rest frame
  $I$-band Tully-Fisher relation compiled from Mathewson, Ford \&
  Buchhorn (1992) (solid points) and from the Ursa-Major Cluster (open
  circles; Verheijen 2001).
}
\label{fig:tf_plot}
\end{figure*}

\subsection{Comparison to Local Populations}
\label{sec:tfrelation}

In Fig~\ref{fig:tf_plot} we show the Tully-Fisher relation for the high
redshift galaxies as compared to previous measurements of the TF in
both low redshift \citep{Pierce92,Verheijen01,MathewsonFord92} and high
redshift \citep{Vogt99} galaxy populations.  Whilst \citet{Vogt99} and
\citet{MathewsonFord92} use optical spectra and \citet{Pierce92} and
\citet{Verheijen01} use H{\sc i} line widths to measure the TF
relation, as \citet{MathewsonFord92} show, with high quality
observations the rotation curve velocity determinations and H{\sc i}
line widths are tightly correlated.  Indeed, the scatter between
$v_{opt}$ and $v_{HI}$ is typically less than 10$\kms$ over a range of
60$\lsim$$v$$\lsim$300 ($\kms$); (Fig.~5 of \citealt{MathewsonFord92})
and therefore it is useful to include the local H{\sc i} sets in our
analysis.

The four galaxies with well defined rotation curves are shown as solid
points.  As can be seen the points suggest a good correlation between
magnitude and rotations speed similar to that seen in the local
Universe.  By assuming the local slope of the TF from \citet{Tully00}
we measure an offset of $M_{B}$=0.41$\pm$0.34mag of brightening in the
rest-frame $B$-band at fixed circular velocity and place a limit of
$<$0.10mag of brightening in the rest-frame $I$-band.  As we discuss
below, the small offset in the I-band TF relation is consistent with
the evolution expected in heirachical galaxy formation models.

In order to plot the remaining two galaxies in Fig~\ref{fig:tf_plot},
we must estimate their circular velocity indirectly.  A\,2390arcB
appears to be a low-mass H{\sc ii} region, and we place a limit of
$\lsim$60$\kms$ on any possible velocity gradient (this is also
comparable to the [O{\sc ii}] emission line FWHM).  Since we are not
able to constrain the inclination angle on this galaxy, we place this
galaxy on the TF relation assuming it is edge on, but also include an
error-bar which indicates a canonical inclination of 60$^{\circ}$.  In
the case of A\,2390arcD, the galaxy is well resolved but does not show
a coherent velocity field.  In this case, we estimate the rotation
speed by assuming the velocity gradient across the galaxy will
eventually develop into regular rotation (it is also possible that this
velocity offsets may develop into a velocity dispersion in an
elliptical galaxy).  The error-bars on the rotation of this galaxy then
represent the uncertainty as to whether this galaxy is relaxed or
merging, as well as an inclination correction assuming an inclination
of 60$^{\circ}$ (as above).

This procedure allows us to show that the position of these point in
the TF relation is reasonable given their optical and near-infrared
brightness, however, the uncertainties in these rotation speed
estimates are clearly large.  These two galaxies should therefore be
interpreted with caution.  Nevertheless, if these two galaxies are
included in the analysis we derive an offset of $\Delta
M_{B}$=0.51$\pm$0.28 and $\Delta$M$_{I}<$0.10.

\begin{figure}
  \centerline{\psfig{file=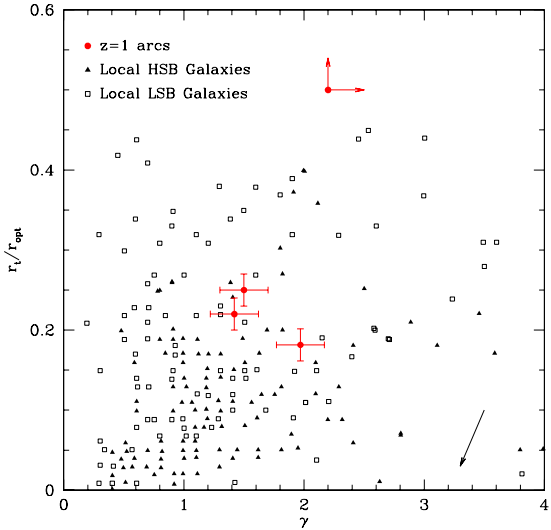,width=3.0in,angle=0}}
\caption{The distribution of $r_{t}/r_{opt}$ against $\gamma$ for the arcs 
  with well defined rotation curves in our sample as compared to local
  ($z\lsim$0.03) spiral galaxies by Courteau et al.\ (1997).  The arrow
  in the lower right hand corner shows the maximum affect the seeing
  FWHM can have on the shape parameters.  The rotation curve from the
  arc in RGB1745+398 is uncertain because the IFU field of view only
  just reaches the turn over and we therefore place lower limits on on
  $r_{t}/r_{opt}$ and $\gamma$ for this galaxy.  The $z$=1 galaxies in
  our sample have source-frame brightnesses consistent with HSB
  galaxies, yet the rotation curve shapes are more consistent with
  low-surface brightness galaxies in the local Universe: if the maximum
  effect the seeing can have on the shape parameters is taken into
  account, then only about 25\% of local HSB galaxies have similarly
  shaped rotation curves.}
\label{fig:rotcurve_shapes}
\end{figure}

A key gain from studying lensed galaxies is the level of detail which
we can extract from their rotation curves.  For the four galaxies with
well defined rotation curves, we can also compare the shapes of the
rotation curves with those of local (z$\lsim$0.03) late-type spiral
galaxies by \citet{Courteau97}.  This sample is dominated by late-type
(Sb-Sc) spirals and therefore should provide a reasonable comparison
sample.  We fit the observed source-plane 1-D rotation curves in our
sample with the {\sc arctan} and multi-fit parameter models from
\citet{Courteau97}.  Using a $\chi^2$ method the observed rotation
curves are best fit with models with high values of $\gamma$ and
r$_{t}$/r$_{opt}$; (see Fig.~\ref{fig:rotcurve_shapes}).  In this
formalism, $\gamma$ governs the degree of sharpness of the turn over of
the rotation curve, r$_{t}$ describes the transition radius between the
rising and flattening of the curve and r$_{opt}$ is the radius
enclosing 83\% of the light in the source plane photometry.  We note
that the rotation curve from the arc in RGB1745+398 is uncertain
because the IFU field of view only just reaches the turn over, however,
the data is sufficient to place a lower limit on $r_{t}/r_{opt}$ and
$\gamma$.  Such models indicate slowly rising rotation curves and
usually favor fainter (or low-surface brightness) galaxies, yet the
source frame surface brightnesses of the four galaxies are consistent
with high surface brightness (HSB) galaxies ($\mu_{b,o}$=21.5, 21.2,
20.6 and 21.0 for A\,2390arcA, Cl2236-04arc, A\,2218\,arc\#289 and
RGB\,1745+398arc respectively).  In order to investigate how sensitive
the shape of the rotation curve is to the seeing we perform two checks:
firstly, we convolve the rotation curve with a further 0.7$''$ seeing
(transformed to the source plane; Fig.~\ref{fig:rot_curves}) and refit
the rotation curve.  Whilst the results vary between the galaxy
rotation curves, the overall effect is only modest at best.  Typically,
we find that the inner rotation curve flattens the transition radius by
$\Delta r_{t}/\Delta r_{opt}$=0.05--0.08, and $\Delta r_{t}/\Delta
r_{opt}$=0.05--0.09, and $\gamma$=0.18 in the {\sc arctan} and
multi-parameter fits respectively.  Secondly, we generate mock rotation
curves for both the {\sc arctan} and multi-parameter fits with low
values of $r_{t}$ and $\gamma$ and convolve these with 0.7$''$ seeing.
For the values of $r_{t}/r_{opt}$ and $\gamma$ in the rotation curves
in our data, we find that they could be affected by upto $\Delta
r_{t}/r_{opt}=$0.06 (arctan), $\Delta r_{t}/r_{opt}=$0.5 (multi-param)
and $\Delta\gamma=$0.09.  In Fig~\ref{fig:rotcurve_shapes} we therefore
show arrow which shows the maximum effect which the seeing could have
on the measured shapes of the rotation curves.  Whilst this contributes
to the shape parameters we find that in the local galaxy population,
only about 25\% of galaxies have similarly shaped rotation curves.
This could be an indication that the inner parts of the galaxy mass
distribution are less baryonically dominated than in local spirals and
could be an indication that the bulge-to-disk mass ratio of these high
redshift galaxies are lower than those of most present day galaxies.

\section{Discussion and Conclusions}\label{sec:discussion}

In this study we have mapped the two dimensional velocity fields of six
$z\sim1$ gravitationally lensed galaxies using IFU spectroscopy.  Using
detailed mass models for the clusters we have reconstructed the source
morphologies of these galaxies.  The typical boost in magnitude for
these galaxies is $\Delta m=2$, which means the typical (unlensed)
$R$-band magnitude of our sample of m$_{R}\gsim22$.  Moreover, the gain
in spatial magnification for a typical galaxy in this paper means that
0.6$''$ on the sky corresponds to only $\sim$0.5\,kpc in the source
frame which allows us to spatially resolve the dynamics of these
galaxies on much smaller scales than otherwise possible.  Our IFU
spectroscopy of these distant galaxies allows us to test whether these
high redshift galaxies have regular disk-like kinematics, or whether
velocity offsets come from merging components or other dynamical
disturbance.

With a small sample of high redshift galaxies which show bi-symmetric
(disk-like) kinematics we can measure the evolution of the offset in
the TF relation under the assumption that the slope remains fixed.  In
Fig.~\ref{fig:tf_plot}, we compare the rotation velocities and source
brightness of the arcs which resemble galaxy disks with that of local
and other high redshift galaxies in rest frame $B$ and $I$-bands.  We
also include the A\,2390arcB and A\,2390arcD in this plot.  Whilst
these two galaxies do not resemble rotating disks and therefore cannot
be compared directly to local disks, we show the likely range of
equivalent circular velocities based on the upper limit to the velocity
dispersion (A\,2390arcB) and assuming that the two components are on a
merging orbit (A\,2390arcD).  These data points are of course highly
uncertain, not least because we have no way to estimate their
inclinations on the sky.

Whilst previous high redshift studies have concentrated on the
rest-frame $B$-band, rest-frame $I$-band observations provide a more
rigorous test of evolution of the TF relation since the corrections for
dust and on-going star formation are much smaller at longer wavelengths
\citep{Conselice05}.  The rest frame $I$-band TF therefore gives a
clearer indication of the true stellar luminosity and hence the ratio
of stellar mass to total halo mass.  The position of the galaxies on
the TF relation in both the $B$ and $I$-bands shows good agreement with
local data \citep{Pierce92,MathewsonFord92,Haynes99,Verheijen01}.  This
data is in agreement with existing intermediate and high redshift
studies, \citep{Vogt97,Vogt99,Vogt02,
  Ziegler02,Milvang-Jensen03,Barden03, Boehm03,Bamford05,Conselice05}.
Overall, our observations suggest a 0.5$\pm$0.3mag of brightening in
the $B$-band TF from the local ($z=0$) correlation, whilst in the
$I$-band we place a limit of $<0.10$mag between our $z=1$ sample and the
local $z=0$ correlation.

The theoretical evolution of the $I$-band TF relation from hierarchical
models of galaxy formation from \citet{Cole2k} predict that for any
given disk circular velocity, the $I$ band luminosity should decrease
by $\sim0.1$ magnitudes from $z=0$ to $z=1$, whilst in the $B$-band
such models predict an increase in luminosity of $\sim0.5$ magnitude
for the same redshift change.  It is useful to compare this prediction
with the prediction for a simple ``classical'' galaxy formation.  We
consider a model in which all of the galaxy's mass were already in
place at $z=1$, but only half of the stars have yet formed
\citep{Eggen62,Hopkins00}. This toy model produces an evolution in
luminosity of $\sim0.7$ magnitudes from $z=0$ to $z=1$, during which
the asymptotic circular velocity remains constant.  The small offset of
our galaxies from the local $I$-band TF relation suggests a preference
for hierarchical rather than the ``classical'' formation model.
Furthermore, the 0.51$\pm$0.28 magnitude evolution in the rest-frame
$B$-band is in line with the increased star formation activity at $z=1$
suggested by studies of the B-band luminosity function
\citep{Giallongo05}.

Clearly, it would be dangerous to draw far reaching conclusions from a
small number of high-redshift galaxies. Rather this study should be
view as a companion to studies of a larger samples of field galaxies
\citep[e.g.][]{Flores04b}.  The advantage of our lensed study is that
we are able to resolve rotation curves in detail and clearly identify
the asymptotic velocity of the rotation curve rather than fitting a
model rotation curve (based on local galaxies) convolved with ground
based seeing.  In contrast to many ground based studies our rotation
measurements extend to large radii allowing asymptotic velocities to be
directly measured.

The present sample is just large enough to place constraints on the
evolution of the TF relation, but a significantly larger sample of
galaxies are accessible to observation using the gravitational
telescope technique we have illustrated (there are currently a further
$\gsim$10 highly magnified and spatially resolved z$\sim$1 giant arcs
in the literature which are suitable for this study; \citealt{Sand05}).

However, the principle gain from studying distant galaxies using
gravitational telescopes is to study the internal properties of the
galaxies in detail. In principle, the boost from gravitational lensing
allows us to achieve an angular resolution to galaxies observed with
classical techniques at $z\sim$0.1 (e.g. for a typical $z\sim$1
gravitationally lensed galaxy in this sample, the spatial sampling of a
galaxy is a factor of three smaller than unlensed galaxies at the same
redshift; by contrast this spatial sampling is matched to unlensed
galaxies at $z\sim$0.15).  This increase in resolution allows us to
study the dynamics, metal abundance and star formation of $z\sim$1
galaxies in detail and hence address {\it why} the global properties of
distant galaxies differ from their local counter parts.

In this paper, we restrict our attention to the dynamics of the target
systems.  For those galaxies with well defined rotation curves we can
use the gravitational magnification to compare the shapes of the
rotation curves with those of local (z$\lsim$0.03) spiral galaxies
observed by \citet{Courteau97}.  We fit the observed source-plane 1-D
rotation curve with the {\sc arctan} and multi-fit parameter models
from \citet{Courteau97}.  Using a $\chi^2$ fit the observed rotation
curves are best fit with models with high values of $\gamma$ and
r$_{t}$/r$_{opt}$.  The fits from these models show that the high
redshift targets have slowly rising rising rotation curves compared to
typical local galaxies.  In the local Universe, such rotation curves
are associated with fainter, low-surface brightness galaxies, yet the
source frame surface brightnesses of the four galaxies are consistent
with high surface brightness (HSB) galaxies.  In the local galaxy
population, only about 25\% of galaxies have similarly shaped rotation
curves and could be an indication that the bulge masses of these high
redshift galaxies is somewhat lower than those at the present day.

The dynamics of two of the galaxies in our sample do not resemble
galaxy disks.  The first, A\,2390arcB appears to be a highly magnified
H{\sc ii} galaxy, with a velocity dispersion of $\sim60\kms$ and a
(source frame) radius of $\sim$2.5\,kpc (FWHM).  The other galaxy in
our sample which does not appear to have stable disk kinematics is
A\,2390arcD.  This galaxy has a disturbed morphology and appears to
have several components.  When combined with the velocity field from
the [O{\sc ii}] emission, we find some evidence for rotation or
interaction. The large [O{\sc ii}] line widths ($\gsim300\kms$ FWHM)
may originate from either outflowing material driven by AGN activity,
or as a result of two interacting galaxies.  This second scenario is
supported by the fact that we also identify a bright knot, or companion
offset by 5\,kpc and 480$\pm$60$\kms$ in projection; the most likely
interpretation of the system is that this galaxy has recently undergone
a tidal interaction/merger which has produced the disturbed and complex
morphology.  The selection criteria for our analysis was simply that
the galaxies must be highly amplified, and therefore it is interesting
to note that although we only have a small sample, of the six galaxies
we have studied, two do not have disk-like kinematics or morphologies,
giving some indication as to the mix of stable disk-like galaxies to
currently assembling galaxies at $z\sim$1.

The next step in this study is to complement these observations with
similar observations in the near-infrared.  At $z\sim$1, H$\beta$,
[O{\sc iii}] and H$\alpha$ are redshifted to beyond 1$\mu$m and until
recently have been inaccessible to Integral Field Spectroscopy (IFS).
However, recent developments in designing efficient image slicing IFU's
allow us to probe beyond 1$\mu$m, and therefore we can combine the
rest-frame optical emission lines to probe the distribution of
reddening through the well studied $R_{23}$ index; $R_{23}$=([O{\sc
  ii}]+[O{\sc iii}])/H$\beta$ \citep{Zaritsky94}, as well as the
reddening corrected star formation rates through the H$\alpha$:H$\beta$
decrement.  For those galaxies which we have identified as disk-like at
$z\sim$1, such observations will give unique insight into the processes
of disk galaxy formation, and give important constraints on the
assembly on the disks of present day galaxies.

The gain in spatial resolution which we have achieved by using a
gravitational lens to magnify and stretch distant galaxies demonstrates
the science that will soon be possible with Adaptive Optics Integral
Field Spectroscopy (AO-IFS) on (non-lensed) galaxies at $z=1$ (e.g.\,
OSIRIS on Keck, NIFS on Gemini and SINFONI on the VLT). These studies
will probe the structure of young galaxies on 100milli-arcsecond (mas)
scales, although long exposures will be required to compensate for the
lack of lensing amplification.  For a non-lensed source at $z=1$, this
corresponds to 600pc. However, the greatest gains will come from
combining AO-IFS with the powerful gravitational lensing technique. For
a lensed galaxy with magnification factor of ten in flux (a factor of
three in linear scale), the effective diffraction limit of the
telescope is reduced by a factor three and 100mas corresponds to only
200pc in the source frame of the galaxy, sufficient to resolve the
star-formation, kinematic and chemical properties of individual H{\sc
  ii} regions (which typically have sizes less than 400pc in local
spiral galaxies; \citealt{Gonzalez97}). We look forward to undertaking
these studies in the coming years.

\section*{acknowledgments}
We are very grateful to the anonymous referee for providing a number of
suggestions which significantly improved the content, layout and
clarity of this paper.  We would like to thank Kari Nilsson for
allowing us to use his $BVRI$ imaging data of RGB\,1745+398, and
Alastair Edge, John Lucey, Chris Simpson and Russel Smith for useful
discussions.  We would also like to thank the Gemini-North staff who
observed our targets in queue mode in 2003A.  GPS thanks Phil Marshall
and Keren Sharon for assistance with the Keck/LRIS observations and
David Sand for assistance with the Hale/WIRC observations.  GPS also
acknowledges the Caltech Optical Observatories TAC for enthusiastically
supporting his program of galaxy cluster observations.  AMS
acknowledges support from a PPARC Fellowship, RGB acknowledges a PPARC
Senior Fellowship, GPS acknowledges support from Caltech and a Royal
Society University Research Fellowship, IRS acknowledges support from
the Royal Society and JPK thanks support from CNRS and Caltech for
their support.

\bibliographystyle{apj}
\bibliography{/home/ams/Projects/ref}

\end{document}